\documentclass[fleqn,10pt]{wlscirep}
\usepackage[utf8]{inputenc}
\usepackage[T1]{fontenc}
\usepackage{comment}
\usepackage{array, makecell}
\usepackage{makeidx}
\usepackage{nicefrac}
\usepackage{tikz}

\usepackage[titletoc]{appendix}
\usepackage{amsmath,amsthm,amssymb}
\usepackage{graphicx,amsfonts,amsbsy}
\usepackage{xcolor}
\usepackage{cite}
\usepackage[percent]{overpic}

\usepackage{siunitx}
\usepackage{hyperref}
\usepackage[capitalise]{cleveref}
\usepackage{graphicx}
\usepackage{dcolumn}
\usepackage{bm}
\usepackage{subcaption}
\usepackage{mlmodern} 
\newcommand{\del}{\partial} 
\newcommand{\field}{\vec{\mathcal{H}}}

\renewcommand{\vec}{\boldsymbol}
\usepackage{mathtools}
\usepackage{booktabs}
\usepackage{float}
\usepackage{wrapfig}

\title{A proposal for leaky integrate-and-fire neurons by domain walls in antiferromagnetic insulators}
\author[1,$\dagger$,*]{Verena Brehm}%
\author[1,$\dagger$]{Johannes W. Austefjord}%
\author[2]{Serban Lepadatu}%
\author[1]{Alireza Qaiumzadeh}%

\affil[1]{Center for Quantum Spintronics, Department of Physics, Norwegian University of Science and Technology, 7491 Trondheim, Norway}
\affil[2]{Jeremiah Horrocks Institute for Mathematics, Physics and Astronomy, University of Central Lancashire, Preston, PR1 2HE, United Kingdom}

\affil[*]{verena.j.brehm@ntnu.no}
\affil[$\dagger$]{these authors contributed equally to this work}

\begin{abstract}
Brain-inspired neuromorphic computing is a promising path towards next generation analogue computers that are fundamentally different compared to the conventional von Neumann architecture.
One model for neuromorphic computing that can mimic the human brain behavior are spiking neural networks (SNNs), of which one of the most successful is the leaky integrate-and-fire (LIF) model. 
Since conventional complementary metal-oxide-semiconductor (CMOS) devices are not meant for modelling neural networks and are energy inefficient in network applications, recently the focus shifted towards spintronic-based neural networks. 
In this work, using the advantage of antiferromagnetic insulators, we propose a non-volatile magnonic neuron that could be the building block of a LIF spiking neuronal network. In our proposal, an antiferromagnetic domain wall in the presence of a magnetic anisotropy gradient mimics a biological neuron with leaky, integrating, and firing properties. This single neuron is controlled by polarized antiferromagnetic magnons, activated by either a magnetic field pulse or a spin transfer torque mechanism, and has properties similar to biological neurons, namely latency, refraction, bursting and inhibition.
We argue that this proposed single neuron, based on antiferromagnetic domain walls, is faster and has more functionalities compared to previously proposed neurons based on ferromagnetic systems.

\end{abstract}

\begin{document}

\flushbottom
\maketitle
\thispagestyle{empty}

\section{Introduction}\label{section:introduction}
Modern electronic digital computers are designed based on the socalled von Neumann computing architecture. They rely on central processing units (CPU), built upon complementary metal-oxide-semiconductor (CMOS) transistors \cite{mahmoud_introduction_2020}. In contrast to that, inspired by the human brain and its complex neural network, novel energy efficient analogue computing architectures with strongly interconnected processing elements have been proposed that lead to the emerging technology of  neuromorphic computing and engineering \cite{Christensen_2022,Low-Power-Computing,Spike-based-Neuromorphic,Communication-consumes}. 

The conventional CPU-based von Neumann computing architecture is faster than the current state of the art neuromorphic computing, but the latter potentially can solve computationally intensive tasks, like speech and character recognition, while offers a more energy efficient data processing \cite{roadmap}.
To achieve even higher energy efficiency as well as faster data processing in neuromorphic computing architecture, it was proposed very recently that neuromorphic principles may be implemented in spintronic-based nanodevices. 
This leads to the emerging field of the \textit{neuromorphic spintronics} \cite{grollier_neuromorphic_2020}.
In spintronic-based nanotechnology, the intrinsic spin angular momentum of electrons, rather than their charge, may be used for data storage and processing. The magnetic insulators that host magnons and various topological magnetic textures are key ingredients for efficient data processing and information storage \cite{BRATAAS20201}. Consequently, ubiquitous Joule heating arising from electron scatterings in metals and semiconductors is avoided in insulators. 
Consequently, several ferromagnetic-based  LIF neurons for SNN networks have already been proposed \cite{IEEEreviewFM,IEEEbringerMTJ,BringerAnisoGradient}. However, recent theoretical and experimental advances in spintronics have shown that antiferromagnetic (AFM) systems have even much more advantages compared to their ferromagnetic (FM) counterparts \cite{Antiferromagnetic_spintronics}. 
The absence of parasitic stray fields, operating at THz frequencies in contrast to GHz in FM systems, existence of opposite chiralities of magnons, and the abundance of room temperature AFM materials in nature, make AFM-based spintronics as a highly promising candidate for the hardware implementation of the next generation of ultrafast, low-energy-cost, and miniaturized non-volatile neuromorphic chips. \cite{Neuromorphic_computing, PhysRevLett.125.207202, Artificial_neurons,Bindal_2021}.

Spiking neural networks (SNNs) are a class of neuromorphic computing architecture that mimic human neural networks \cite{MAASS19971659}. One of the most successful spiking neural network models is the leaky integrate-and-fire (LIF) model \cite{gerstner_spiking_2002}. This model resembles the spiking behavior of a neuron at the onset of critical accumulating stimuli and its slow decay to the equilibrium state until the next spike \cite{STEIN1965173}. LIF may be used as the building block of neuromorphic chips \cite{https://doi.org/10.1002/adfm.201604740}.

In this paper, we propose a non-volatile AFM-based single neuron with leaky integrate-and-fire properties  that may be the building block for a LIF spiking neural network. The state of this neuron is encoded in the position of a domain wall (DW), which is controlled by AFM magnons. Leaky behavior is ensured by a nonuniform magnetic anisotropy profile, while there is no standby leakage in the neuron.

\section{Theory of Neural Networks}\label{SecModelTheory}
In this section, we briefly summarize the key elements and ingredients of SNN and LIF single neuron models. In the next sections, we show our proposed single neuron has similar characteristic.

\subsection{Spiking Neural Networks} \label{sec:snn}
A SNN takes the inspiration of human brain activity into computer science one step further than other models of artificial neural networks, like feedforward neural networks \cite{maass_networks_1997}. Information in this model is encoded as spike trains; c.f., binary information coding, used in conventional computers. The network has an explicit time dependency and the system is event-driven. 
\begin{figure}[t]
    \centering
    \includegraphics[width = 0.7\linewidth]{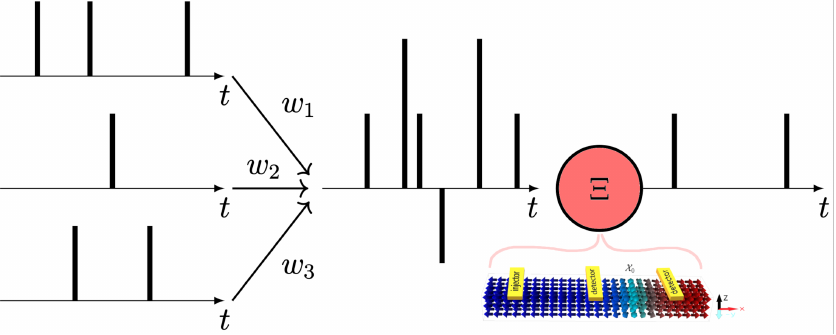}
    \caption{Schematic of a spiking neural network \cite{jang_spiking_2021}. A LIF neuron $\Xi$ receives input spikes from several presynaptic neurons. In the present work, we model $\Xi$ by an AFM DW. The spike trains are multiplied by weights $w_i$ and merged before they get sent into $\Xi$. A non-linear function determines whether the neuron should fire as a consequence of stimuli from its synapses.
    \label{fig:SNN}}
\end{figure}
We first give a brief mathematical description of the SNN model. A generic spiking neuron $\Xi$ is represented in \cref{fig:SNN}. 
Let $V$ be a finite set of spiking neurons, connected by a set of $E \subseteq V \times V$ synapses. For each synapse $\langle i, j \rangle \in E$ between presynaptic neuron $j$ and postsynaptic neuron $i$ there is associated a response function $\epsilon_{ij}$ and a weight $w_{ij}$. The state variable of $i^{\rm{th}}$ neuron, $u_i(t)$, is then given by \cite{gerstner_spiking_2002,maass_networks_1997},  
\begin{equation}\label{eq:SRM0}
	u_i(t)  =\delta(t - t_i^{(f)}) + \sum_j \sum_f w_{ij}\epsilon_{ij}(t - t_j^{(f)}) + u_0.
\end{equation}
Here $u_0$ is the equilibrium potential, i.e. the value of $u_i(t)$ when no stimuli has affected the neuron and $t_j^{(f)}$ indicates the firing times, where $f$ is the label of each spike. In general the firing time \(t = t_i^{(f)}\) of a neuron $i$ is set when $u_i(t)$ reaches a threshold value \(u_\text{threshold}\),
\begin{equation}\label{eq:threshold}
\begin{aligned}
	&u_i(t) = u_\text{threshold} \quad \wedge \quad \text{sgn}(u_i(t) - u_0) \frac{du_i(t)}{dt} > 0 \\ &\Longrightarrow t = t_i^{(f)},
\end{aligned}
\end{equation}
where $\text{sgn}(x)$ is the sign function and $\epsilon_{ij}(t - t_j^{(f)})$ determines the response for postsynaptic neuron $i$ from stimuli from presynaptic neuron $j$. Once a spike is initiated, $u_i(t)$ is immediately reset to $u_0$.
Equation \eqref{eq:SRM0} can therefore be used to model a human neuron: after the action potential in a neuron has been raised and neurotransmitters have been transferred, it relaxes back to its ground state until the next activation happens.  

It is worth noting that Eq. (\ref{eq:SRM0}) assumes no time delay as signals travel the synapses. This could easily be added with a delay time for each synapse \cite{bohte_error-backpropagation_2002}.

\subsection{Leaky Integrate-and-Fire Neurons}

\begin{figure}
    \centering
    \includegraphics[width=0.4\linewidth]{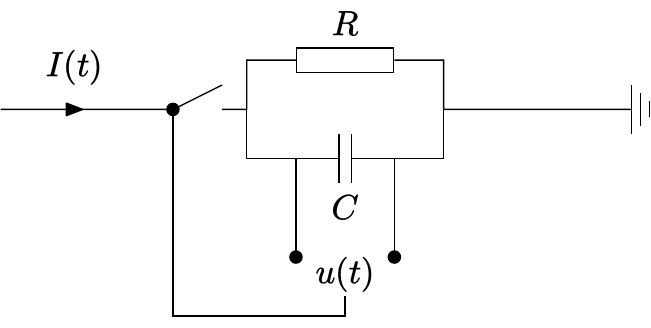}
    \caption{Leaky integrate-and-fire circuit \cite{gerstner_spiking_2002}. A capacitor, $C$, and a resistor, $R$, are connected in parallel. The voltage over the capacitor $u(t)$ integrates the current input, while it leaks to ground. When $u(t)$ reaches a threshold value, a switch controlling the input wire is flipped, stopping new currents into the system for a refractory period. During the refractory period charge is completely depleted from the capacitor.}
    \label{fig:LIF}
\end{figure}
The rather general \cref{eq:SRM0} can be used to model a variety of neuron models.
LIF models are one of the most prominent neuron types \cite{gerstner_spiking_2002}. It can be modelled by a resistor–capacitor circuit ($RC$) circuit as shown in \cref{fig:LIF}. The neuron voltage corresponds to the capacitor voltage $u_i(t)$. The LIF model is described by a differential equation,
\begin{equation}\label{eq:LIF}
	\tau\frac{du_i}{dt} = -u_i(t) + RI_i(t),
\end{equation}
where $\tau = RC$ is the time constant of the $RC$ circuit, and $R$ and $C$ are the resistance and capacitance of the resistor and capacitor, respectively. 
The incoming current $I_i(t)$ is, 
\begin{equation}
	I_i(t) = \sum_j w_{ij} \sum_f \delta (t - t_j^{(f)}).
\end{equation}
The weights $w_{ij}$ determine the connection strength from presynaptic neuron $j$ to postsynaptic neuron $i$. The sum $\sum_f$ is over all presynaptic spike times $(f)$.

The purpose of the LIF model is to describe how the spiking neuron $\Xi$ behaves as a function of external stimuli, or captures the dynamics of the $\epsilon_{ij}$ response function in Eq. \eqref{eq:SRM0}. The LIF model has a memory of previous inputs $I_i(t)$, stored on the capacitor. The resistor ensures that this memory only is short term. As before, a spike is fired when $u_i(t)$ reaches a threshold value by \cref{eq:threshold}.  A generalization to a non-linear leaky integrate-and-fire model gives
\begin{equation}\label{eq:general_LIF}
	\tau \frac{du_i}{dt} = F(u_i) + G(u_i)I_i(t),
\end{equation}
where the functions $F(u_i)$ and $G(u_i)$ are arbitrary functions. It is worth noting that \cref{eq:SRM0} describes $u_i(t)$ as a function of time since the last input, while Eqs. \eqref{eq:LIF} and \eqref{eq:general_LIF} are implicit equations.

\section{Non-volatile Spintronic-Based LIF Neurons \label{Sec:proposal}} 
In this section, we introduce our proposal of a non-volatile LIF neuron, implemented with a magnetic DW in an AFM insulator with orthorhombic (or biaxial) magnetic symmetry. Although, for computational convenience, we have chosen toy model parameters, see Table \ref{tab:coefficients}, it can be shown that the functionality of the proposed AFM-based neuron is robust against specific material parameters or different system sizes and is scalable by tuning the excitation amplitude and duration. In addition to showing the scalability and robustness of our results, we present the result of micromagnetic simulations with material parameters of hematite in \cref{AppendixHematite}.

\subsection{AFM Model \label{AFM-model}} 
We consider a generic two-sublattice AFM insulator nanoribbon, with orthorhombic magnetic structure, modelled by the following potential-energy density for each sublattice, 
\begin{align}\label{eq:free_energy_density}
    		\mathcal{U}_i(\vec{m}_i, \nabla \vec{m}_i; \vec{r})= & A (\bm{\nabla} \bm{m}_i)^2+ 4A_\text{h} \bm{m_i} \cdot \bm{m_j} - \mu_0 M_\text{s}   \bm{m}_i \cdot \bm{H}  -K_\text{easy} (\bm{m}_i \cdot \vec{\mathrm{e}}_\text{easy})^2  + K_\text{hard} (\bm{m}_i \cdot \vec{\mathrm{e}}_\text{hard})^2  \\ &+ D \bm{m}_i \cdot (\nabla \times \vec{m}_i ) + \eta_i \frac{D_h}{2} \bm{d}_h\times\bm{m}_j,
\end{align}
where $i\neq j \in \{\text{A, B}\}$ refer to two AFM sublattices. Within a micromagnetic framework \cite{AtomMicroScales,lepadatu_boris_2020}, all magnetic contributions in a unit cell with volume $V_0$ are averaged to a macrospin magnetic moment $\bm{M}$, with a saturation magnetization value $M_s=|\bm{M}|$.  The unit vector of magnetization direction is $\bm{m}=\bm{M}/M_s$. 
$A$ and $A_\text{h}$ parameterize the AFM exchange stiffness and the homogeneous Heisenberg exchange interaction, respectively, $K_\text{easy (hard)} > 0$ parameterizes single ion easy (hard) axis anisotropy energy along the $\vec{\mathrm{e}}_\text{easy (hard)}$ direction, $\vec{H}$ is the applied magnetic field, $D$ is the strength of the inhomogeneous bulk-type Dzyaloshinskii-Morya interaction (DMI) while $D_h$ is the homogeneous DMI along the direction $\bm{d}_h$ with a sublattice-dependent sign $\eta_{A (B)} = +(-)1$.


We assume the AFM insulator supports a rigid magnetic domain wall (DW) that connects two uniform AFM domains, see  \cref{fig:setup}. Within the collective coordinate approximation \cite{PhysRevLett.110.127208}, the position of the DW center is considered as a dynamical variable $\mathcal{X}_\text{DW}$. 
In order to control the equilibrium position of DW center, the spatial profile of the anisotropy energy density $K$ can be tuned by electric field via voltage-controlled magnetic anisotropy (VCMA) effect \cite{rana_towards_2019,VCMA1,VCMA2,VCMA3,VCMA4} or strain-induced magnetic anisotropy \cite{https://doi.org/10.1002/advs.201800356,https://doi.org/10.1002/pssr.201900467, https://doi.org/10.48550/arxiv.2209.04527, PhysRevMaterials.4.094004, Meer_StrainInducedAniso,AnisoProfileStress}. We model a spatially varying anisotropy as,
\begin{equation}\label{eq:K_var} \centering
    K(\vec{x}) = K_0\left[\frac{1}{L_x} \left(x - \mathcal{X}_0\right)^2+ 1\right],
\end{equation}
where $L_x$ is the length of the AFM nanoribbon along the $x$-direction. This magnetic anisotropy profile creates a magnetic potential well along the $x$-direction with a minimum value $K_0$ at  $\mathcal{X}_0$ that can be engineered. The AFM DW is at its minimum energy if the DW center is placed at this minimum $\mathcal{X}_0$. If there is no spatial dependent magnetic anisotropy, the system has translation invariance and AFM DWs have no preferred equilibrium position. In our simulations, without loss of generality, we set $\mathcal{X}_0={2L_x}/{3}$.

The spatial dependent of $K(\bm{x})$ ensures that the AFM DW always relaxes back toward its ground-state position $\mathcal{X}_0$ in the absence of stimuli, giving the neuron a {\it{leaky}} behavior. 
 Due to this anisotropy profile, the system is also {\it{non-volatile}} in the sense that the ground state of the neuron is stable. Therefore there is not much standby leakage power in contrast to common CMOS-based neurons.

\begin{table}
	\centering
	\caption{Numerical parameters used for micromagnetic simulations. The according effective field strength for exchange, easy (hard) anisotropy, and DMI are $\mu_0H_\text{exchange}=\SI{400}{T}$, $\mu_0H_\text{easy (hard)}=\SI{20(10)}{T}$ and $H_\text{DMI}=0$--$\SI{0.25}{T}$, respectively.} 
	\begin{tabular}{@{}lll@{}}
		\toprule
		Quantity           & Value                & Unit                 \\ \midrule
		Length of AFM ($L_x$) & $\SI{500}{}$ & $\si{\nano \meter}$ \\
		Width of AFM ($L_y$) & $\SI{20}{}$ & $\si{\nano \meter}$ \\
		Height of AFM ($L_z$) & $\SI{4}{}$ & $\si{\nano \meter} $\\
		Simulation cell size  & $\SI{4}{}$ & $\si{\nano \meter} $\\
		Inhomogeneous exchange stiffness ($A$)                      & 1                    &         $\si{\pico\joule\per\meter}$ \\
		Homogeneous exchange energy ($A_h$)&  $-200$                    &         $\si{\kilo\joule \per \meter\cubed}$ \\
		Easy-axis anisotropy energy ($K_\text{easy}$)&    $20$                  &            $\si{\kilo\joule \per \meter\cubed}$         \\
		Hard-axis anisotropy energy ($K_\text{hard}$) & $10$ & $\si{\kilo\joule \per \meter\cubed}$ \\
		Characteristic length scale $\left(\lambda_\text{easy} = \sqrt{A/2K_\text{easy}}\right)$ & $7$ & \si{nm} \\
		Characteristic length scale $\left(\lambda_\text{hard} = \sqrt{A/2K_\text{hard}}\right)$ & $5$ & \si{nm} \\
		Saturation magnetization ($M_s$)&      $2.1$                  &                 $\si{\kilo\ampere \per \meter}$ \\ 
		Gilbert damping parameter ($\alpha$)           &   $\SI{0.002}{}$                    &                      1\\ 
		Inhomogeneous bulk DMI           ($D$)            &        $0$--$250$             &                    $\si{\micro\joule \per \meter\squared}$  \\
		Homogeneous DMI            ($D_h$)            &        $2$             &                    $\si{\kilo\joule \per \meter\cubed}$  \\
		Applied magnetic field frequency ($\omega$)&  $62.5$ & $\si{\radian \per \pico\second}$\\
		\bottomrule
	\end{tabular}
	\label{tab:coefficients}
\end{table}
\subsection{AFM DWs as LIF Neurons \label{Sec:proposal1}} 
\begin{figure}
    \centering
    \includegraphics[width=0.8\linewidth]{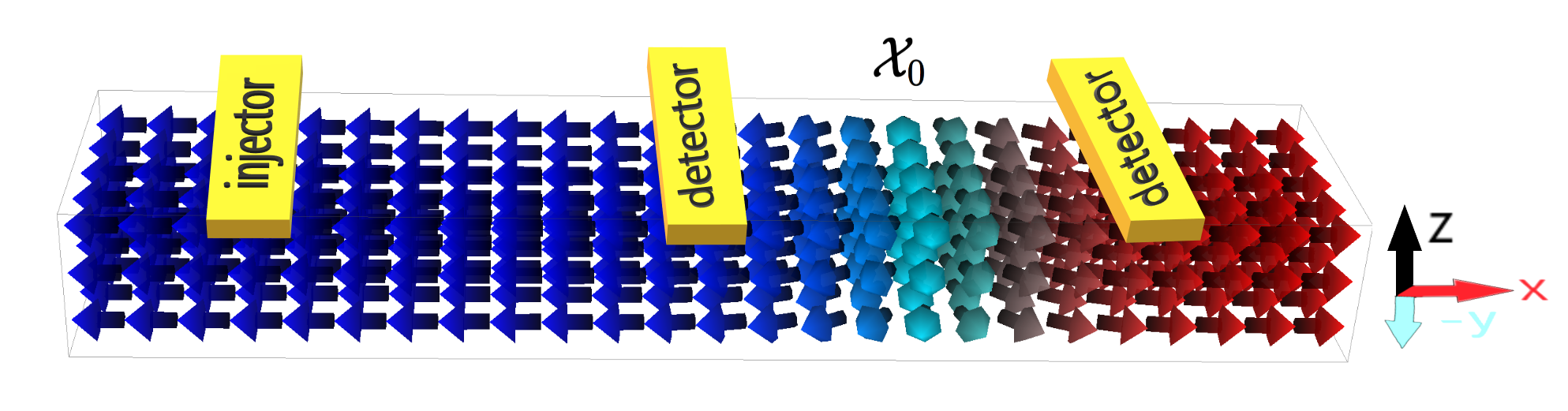}
    \caption{Schematic setup of the AFM-based single neuron proposal in the IP geometry. There are two domains in the AFM stripe, represented by the N\'eel vectors in blue and red. The two domains are connected by a DW texture in turquoise. On top of the AFM stripe, an injector is placed at the left side as a source of magnons and two detectors are placed right and left of the equilibrium position of the DW, the latter shown by $\mathcal{X}_0$.}
    \label{fig:setup}
\end{figure}

Our proposed system is schematically presented in \cref{fig:setup}. It consists of an AFM insulator stripe, an injector (modelling the receptor of a human neuron) that excites magnons in the AFM insulator via either a circularly polarized magnetic field pulse or current-induced (anomalous) spin Hall torque mechanism \cite{tunableLongDistanceKlaui, doi:10.1021/acs.nanolett.8b02114}, and a detector (modelling the transmitter). The detector measures the passing DW via inverse (anomalous) spin Hall effect of the injected spin-pumping signal \cite{ArneSpinPumping, doi:10.1021/acs.nanolett.8b02114, tunableLongDistanceKlaui,https://doi.org/10.48550/arxiv.2211.01195, doi:10.1063/1.4967171}. In a series of neuron networks, this detector or transmitter must be connected to the injector or receptor of the following neuron. At a given set of material parameters and excitation strength, the position of the detector determines the neuron threshold potential. 

AFM DWs are 1D particle-like magnetic solitons that connect two magnetic domains in magnetic materials. It was recently shown that the position of a DW in an AFM insulator is controllable through magnon-DW interactions  \cite{AlirezaHelicityAFMdw}. The position of AFM DW may be used as a state variable for the LIF neuron, $u(t) \longrightarrow \mathcal{X}_\text{DW}$ \cite{agrawal_mimicking_2019}.

In the following, two generic magnetic geometries for possible implementation of LIF neurons are investigated and compared, which we will call in-plane (IP) and out-of-plane (OOP), referring to their magnetic ground-state orientation. 
In order to model these two magnetic states using the potential energy density expression given by \cref{eq:free_energy_density}, we set $\bm{e}_\text{easy}=\hat{e}_x$ and $\bm{e}_\text{hard}=\hat{e}_z$ in IP case, while for OOP, we set $\bm{e}_\text{easy}=\hat{e}_z$ and $\bm{e}_\text{hard}=\hat{e}_x$. Therefore, in the IP geometry, the magnetic ground state lies along the direction of magnon propagation, i.e., the $x$ axis, while in the OOP geometry, the magnetic ground state is normal to the direction of magnon propagation. In both cases, we assume the homogeneous DM vector lies parallel to the hard axis, $\bm{d}_h\parallel\bm{e}_\text{hard}$. 

\subsection{Equation of Motions for AFM Systems}
The dynamics of the normalized sublattice magnetic moments $\vec{m}_{i\in \{\text{A, B}\}}(\bm{r},t)$, in finite temperature, is given by the coupled stochastic Landau-Lifshitz-Gilbert (sLLG) equations, 
\begin{equation}\label{eq:LLG} \centering
	\frac{\del \vec{m}_i}{\del t} = -|\gamma_\text{e}| \mu_0 \vec{m}_i \times (\field_i+\field_i^{th}) + \alpha \vec{m}_i \times \frac{\del \vec{m}_i}{\del t}+\bm{T}(\bm{r},t),
\end{equation}
with the electron gyromagnetic ratio $\gamma_e$, the vacuum permeability $\mu_0$, and the Gilbert damping constant $\alpha$.
The sublattice-dependent effective magnetic field $\field_i = - (\mu_0 M_\text{s})^{-1}\delta U/{\delta \vec{m}_i}$, is given by the functional derivative of 
the total potential energy $U[\vec{m}_\text{A},\vec{m}_\text{B};\vec{r},t] = \int d \bm{r} \sum_i \mathcal{U}_i(\vec{m}_i, \nabla \vec{m}_i; \vec{r},t)$. The current-induced spin transfer torque and magnetic field torque are denoted by $\bm{T}$ in the sLLG equation. $\bm{T}(\bm{r},t)$ is finite only in the injector region and during the excitation period.

Finite temperature dynamics is captured by adding an uncorrelated white noise term in the LLG equations as an effective stochastic magnetic field $\bm{H}^\text{th}$, derived by the fluctuation-dissipation theorem \cite{AtomMicroScales}. It consists of a normalized Gaussian distribution that is scaled with the prefactor $\xi_{th}= \sqrt{\frac{2\alpha k_B T}{\gamma_e \mu_0 M_s V \Delta t}}$, containing the thermal energy $k_B T$, with the Boltzmann constant $k_B$, the cell size volume $V$ and the time step of the simulation $\Delta t$. This prefactor corresponds to $1/\sigma$ in the standard definition of a Gauss distribution. The time step of the simulation is set to $\Delta t=\SI{2}{fs}$ at zero temperature and $\Delta t=\SI{1}{fs}$ at finite temperature.

 In general, spin pumping effect  enhances the local Gilbert damping at injector and detector regions \cite{ArneSpinMixingConductance}. In our simulations, we have ignored this small spin-pumping-induced damping enhancement \cite{PhysRevLett.111.097602}.

To solve coupled sLLG equations for our AFM system, we use the software \textit{Boris Computational spintronics} \cite{lepadatu_boris_2020}.
The list of parameters, used in the micromagnetic simulations, is given in the \cref{tab:coefficients}.

\section{Results}\label{section:results}
In this section, we characterize our proposed non-volatie AFM-based LIF neuron. As we mentioned earlier, AFM DWs are displaced by AFM magnons that can be generated by either magnetic field pulses or by (anomalous) spin Hall torque. 
First, as a proof of concept of AFM-based neurons, we study the interaction between monochromatic magnons, excited by a magnetic field pulse, and AFM DWs at zero temperature. 
Since all-electric control of neurons is the technologically relevant case, in the second part of this section, we show that our proposed single neuron may indeed work by spin Hall torque at finite temperature.

\subsection{Magnon-Induced AFM DW Motion by Magnetic Fields} \label{sec:movingDW}
Magnetic field pulses may excite monochromatic AFM magnons with certain polarizations. It was theoretically shown that these AFM magnons can displace AFM textures in opposite directions depending on their polarizations, values of DMI, and the Gilbert damping parameter  \cite{AlirezaHelicityAFMdw, PhysRevLett.112.147204,PhysRevB.99.054423}.

\begin{table}[htbp]
	\centering
	\caption{Four-stage protocol for magnon-induced DW movement, induced by a transverse magnetic field pulse.}
	\begin{tabular}{@{}lll@{}}
		\toprule
		 Stage   & Magnetic Field Pulse & Polarization \\ \midrule
		Excitation 1 & \thead{$\bm{H}_{IP}(t)=\left(0,H_0 \cos\omega t,H_0\sin \omega t\right)$\\ $\bm{H}_{OOP}(t)=\left(H_0 \cos\omega t,H_0\sin\omega t,0\right)$} & $~~~~\bm{\circlearrowleft}$\\
		Relaxation 1 & $H_0 = 0$ & ~~~~ - \\
		Excitation 2 & \thead{$\bm{H}_{IP}(t)=\left(0,H_0 \sin\omega t,H_0\cos \omega t\right)$\\ $\bm{H}_{OOP}(t)=\left(H_0 \sin\omega t,H_0\cos\omega t,0\right)$}&  $~~~~\bm{\circlearrowright}$\\
		Relaxation 2 & $H_0 = 0$ & ~~~~ - \\
		\bottomrule
	\end{tabular}
	\label{tab:fourstageprotocol}
\end{table}

\begin{figure}
    \centering
    \includegraphics[width=1.0\linewidth]{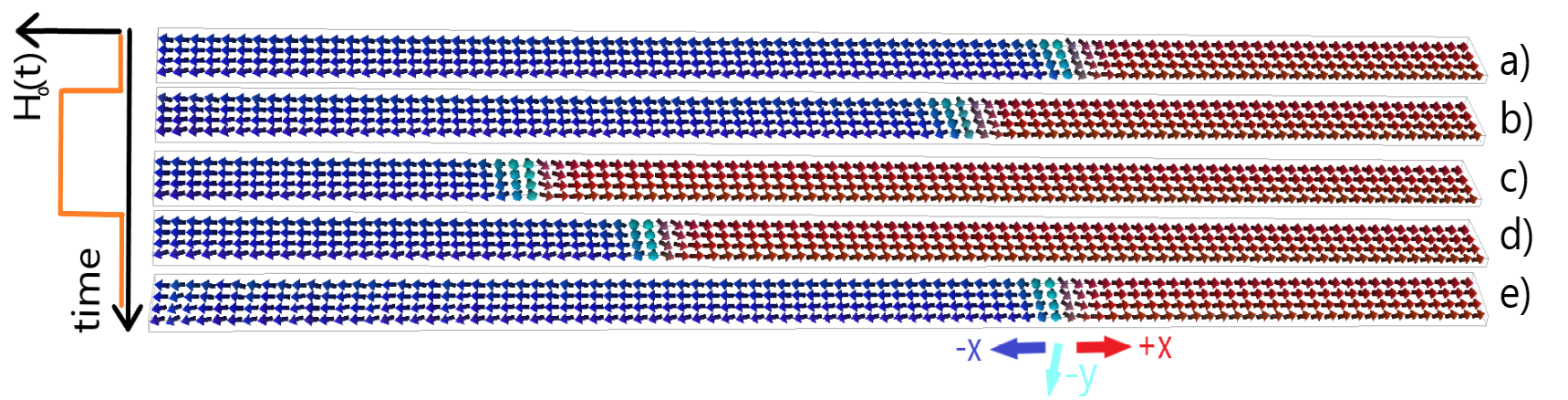}
    \caption{Snapshots of all-magnonic DW motion through an AFM-based neuron in the IP configuration with magnetic field pulse excitation. In (a), the DW is at equilibrium position $\mathcal{X}_{DW}=\mathcal{X}_0$, set by the magnetic anisotropy profile. Once a left-handed magnetic field pulse with strength $H_0$ is turned on, left-handed AFM magnons are excited at the injector. As a result, the DW moves towards the magnon source, panels (b) and (c). After switching the magnetic field off, the DW relaxes back to its equilibrium position, panels (d) and (e). The illustrated movement corresponds to the first excitation stage followed by the first relaxation stage in our protocol. We set $D=\SI{150}{\micro\joule \per \meter\squared}$ in this case.}\label{fig:DWdemo}
\end{figure}

In this part, first, we demonstrate the control of the AFM DW in our setup. To do so, a four-stage protocol is run, see \cref{tab:fourstageprotocol}: In the first excitation stage, a small amplitude transverse magnetic field pulse with circular polarization is applied in the injector region to excite the AFM magnon eigenmodes in the magnetic layer. Afterwards, the magnetic field pulse is turned off and the system may relax back to its ground state in the first relaxation stage. Then, in the second excitation stage, the magnetic field pulse is applied again but with the opposite helicity. Finally, it is turned off again in the second relaxation stage. In \cref{fig:DWdemo}, we present snapshots of magnon-induced AFM DW motion in an IP geometry for one excitation followed by one relaxation stage: while the magnetic field is turned on, the AFM DW travels from its equilibrium position (\cref{fig:DWdemo}a) towards the left (Figs. \ref{fig:DWdemo}b and \ref{fig:DWdemo}c). Once the magnetic field is turned off, it relaxes back toward its equilibrium position (Figs. \ref{fig:DWdemo}d and \ref{fig:DWdemo}e).  AFM-DW motion shows an inertial behavior. When the magnonic forces exerted on the AFM DW vanish, the AFM DW continues to move, until the Gilbert-damping-induced dissipative force stops it and consequently the attractive potential of the magnetic anisotropy pulls it back towards its equilibrium position. This inertial behavior can be seen as a slight overshooting in the DW trajectories presented in Figs. \ref{fig:varyingDMI}, \ref{fig:LeakyIntegrate} and \ref{fig:MoveDWwithTorque} \cite{AFMcollectiveCoord,AFMdwWithSpinWaves}.

By tuning the excitation strength and the distance of the detector from the magnetic anisotropy minimum, one can set the threshold for the firing mechanism. Depending on the strength of the DMI $D$, the DW surface can be tilted. This DMI-induced tilting was also reported in ferromagnetic DWs \cite{PhysRevLett.111.217203}. 

\subsection{Direction and amplitude of the DW displacement}\label{sec:DMIscan}

\begin{figure}
    \centering \hfill
    \begin{subfigure}{0.49\textwidth}
    \centering
    \begin{overpic}[width=1.0\textwidth]{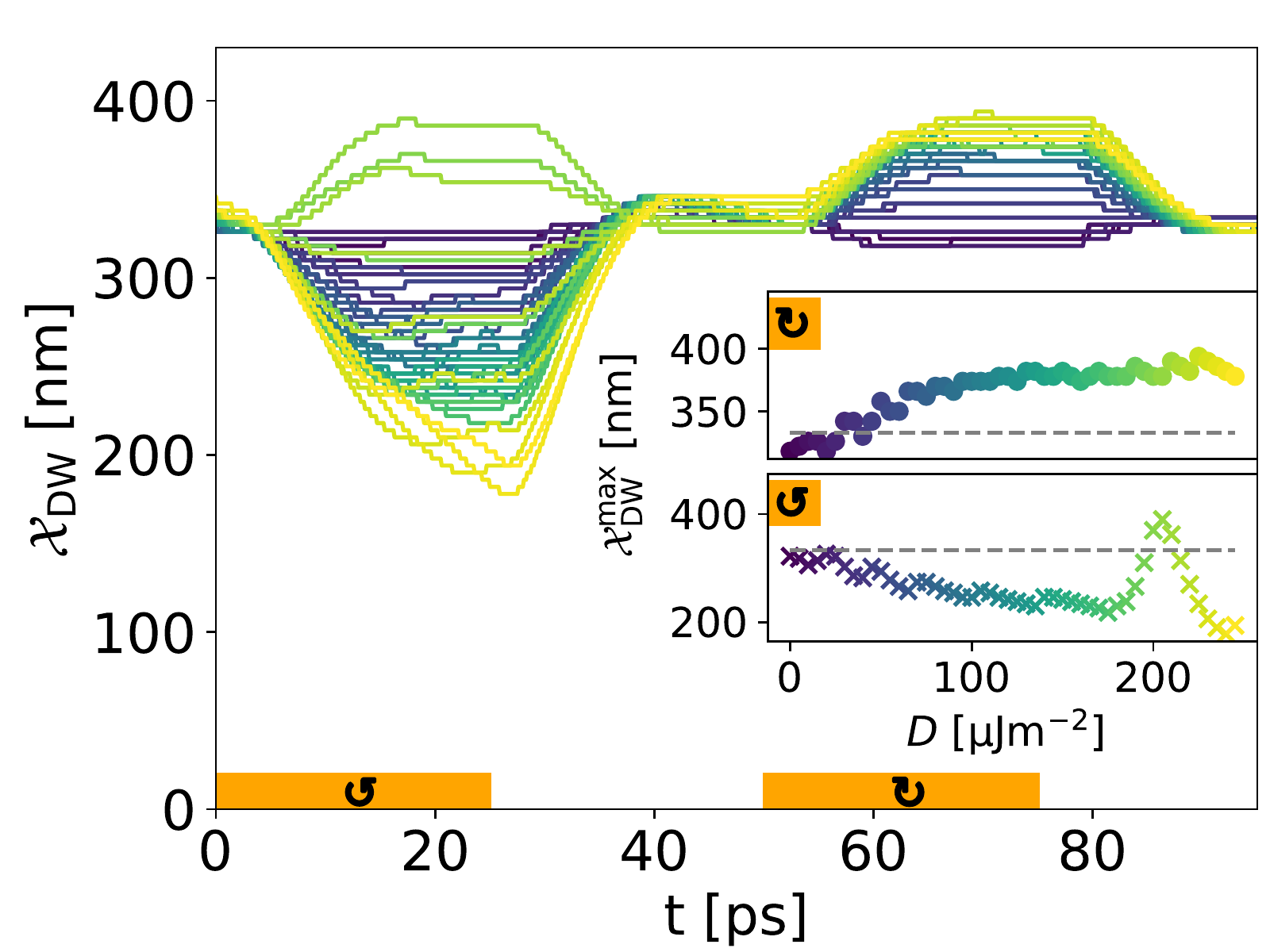}
    \put (10,56) {\small \textcolor{gray}{$\mathcal{X}_0$-}}
    \end{overpic}
    \subcaption{IP geometry \label{fig:VaryingD-IP}}
    \end{subfigure}\hfill%
    \begin{subfigure}{0.49\textwidth}%
    \centering%
    \begin{overpic}[width=1.0\textwidth]{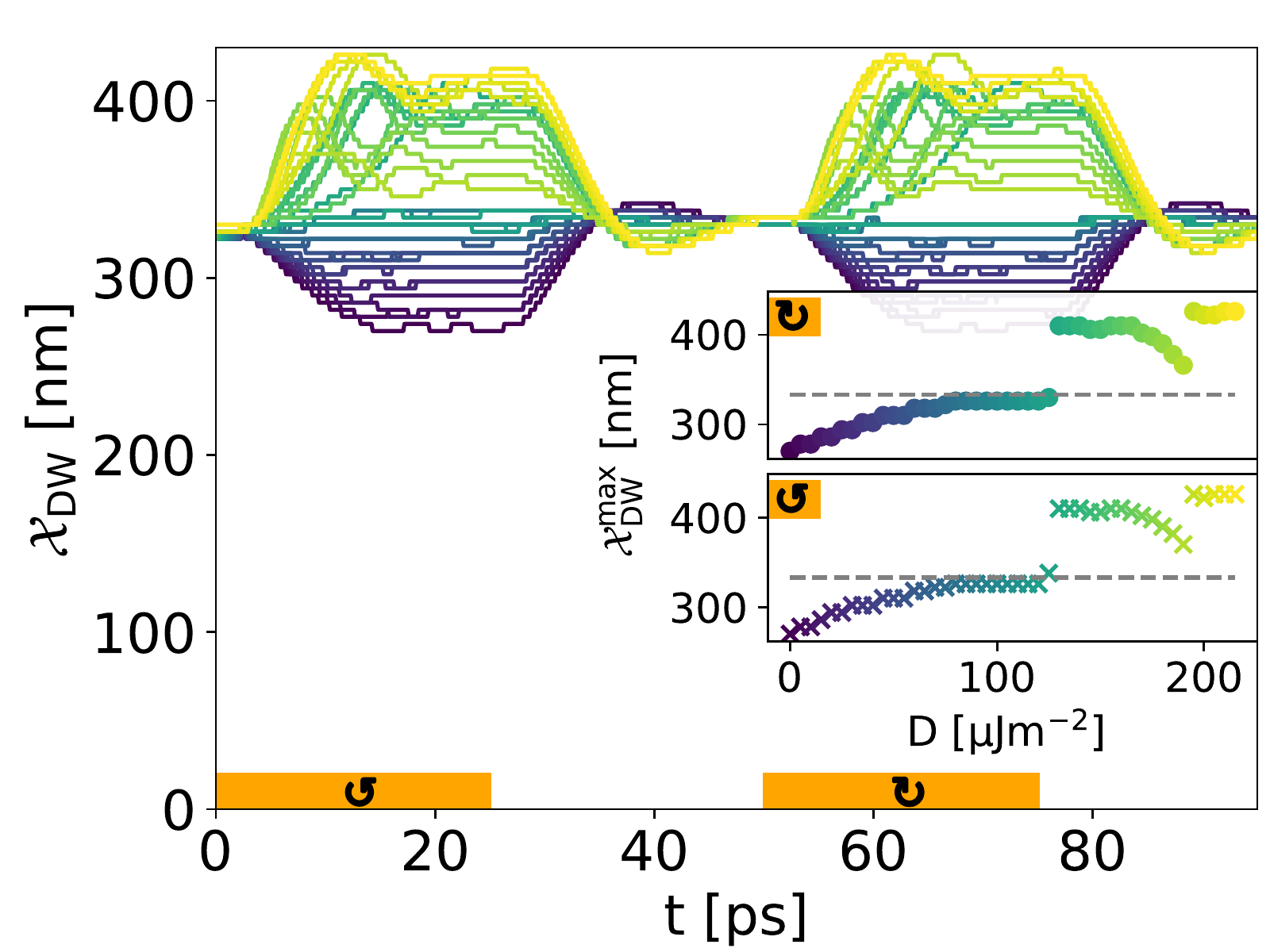}
    \put (10,56) {\small \textcolor{gray}{$\mathcal{X}_0$-}}
    \end{overpic}
    \subcaption{OOP geometry
    \label{fig:VaryingD-OP}}%
    \end{subfigure}\hfill%
    \caption{DMI-dependent all-magnonic AFM DW movement. Left- and right-handed AFM magnons are excited with polarized magnetic field pulses, see the orange area. 
    In the IP geometry (a) the direction and amplitude of the DW motion can be tuned by DMI strength and the chirality of the excited magnons. However, the direction of AFM DW motion in the OOP geometry (b) is independent of the magnon chirality.
    The strength of DMI is encoded by colors, from lowest $D=0$ in blue to highest in yellow, see the insets. 
    In the insets, the maximal displacements of AFM DWs, $\mathcal{X}_\text{DW}^\text{max}$, are shown for each excitation stage (crosses for the first and points for the second excitation stage).}
    \label{fig:varyingDMI}
\end{figure}

In this part, we show that the movement of AFM DWs can be controlled by demand, which makes them more flexible than their ferromagnetic counterpart. Besides the excitation strength (here the magnetic field strength), the magnon polarization, and the inhomogeneous DMI strength have a major impact on the DW displacement. 
In \cref{fig:varyingDMI} the trajectory of the AFM DW center in the IP geometry (\cref{fig:VaryingD-IP}) and OOP (\cref{fig:VaryingD-OP}) is shown during the four-stage protocol, see \cref{tab:fourstageprotocol}. The orange areas in the plots sketch when and where the magnetic field pulse is applied while arrows indicate the helicity of the magnetic field pulse. The color map refers to the strength of the inhomogeneous bulk DMI, starting from dark blue for $D=0$ and increasing over green to yellow for $D=\SI{250}{\micro\joule\per\meter\squared}$ ($D=\SI{200}{\micro\joule\per\meter\squared}$) for the IP (OOP) geometry. Every single line represents one DW trajectory at a given set of parameters. For example, at an intermediate DMI strength, the dark green curve in the IP case (\cref{fig:VaryingD-IP}), the DW moves towards the injector during the first excitation stage ($0$--\SI{25}{ps}), then relaxes back to equilibrium position ($35$--\SI{50}{ps}), and in the second excitation stage with opposite helicity the AFM DW is pushed away from the injector ($50$--\SI{75}{ps}) before relaxing back to the equilibrium position again. 

The first difference between the two cases is the polarization dependency of AFM DW motion. The displacement of an AFM DW in the OOP geometry is insensitive to the polarization of the excited AFM magnons, while the displacement of an AFM DW in the IP case is polarization dependent.
 
Figure \ref{fig:varyingDMI} shows that in the OOP geometry, only the strength of the inhomogeneous DMI determines the direction of the DW motion, but in the IP geometry, both the strength of the inhomogeneous DMI and the chirality of the excited magnons set the direction of AFM DW displacement. 

The amplitude and direction of the maximum displacement of the AFM DW center, $\mathcal{X}_{\rm{DW}}^{\rm{max}}$, show a complicated relation with inhomogeneous DMI strength, see the insets in Figs. \ref{fig:VaryingD-IP} and \ref{fig:VaryingD-OP}.
Recent theoretical studies have shown that, in the presence of an inhomogeneous DMI, several torques and forces act on the AFM DW, and thus the competition between them determines the direction and amplitude of the DW displacement \cite{AlirezaHelicityAFMdw}.

\subsection{LIF Behavior of AFM DWs} \label{sec:pump}
\begin{figure}
    \centering \hfill
    \begin{subfigure}{0.49\textwidth}
    \centering
    \begin{overpic}[width=1.0\textwidth] {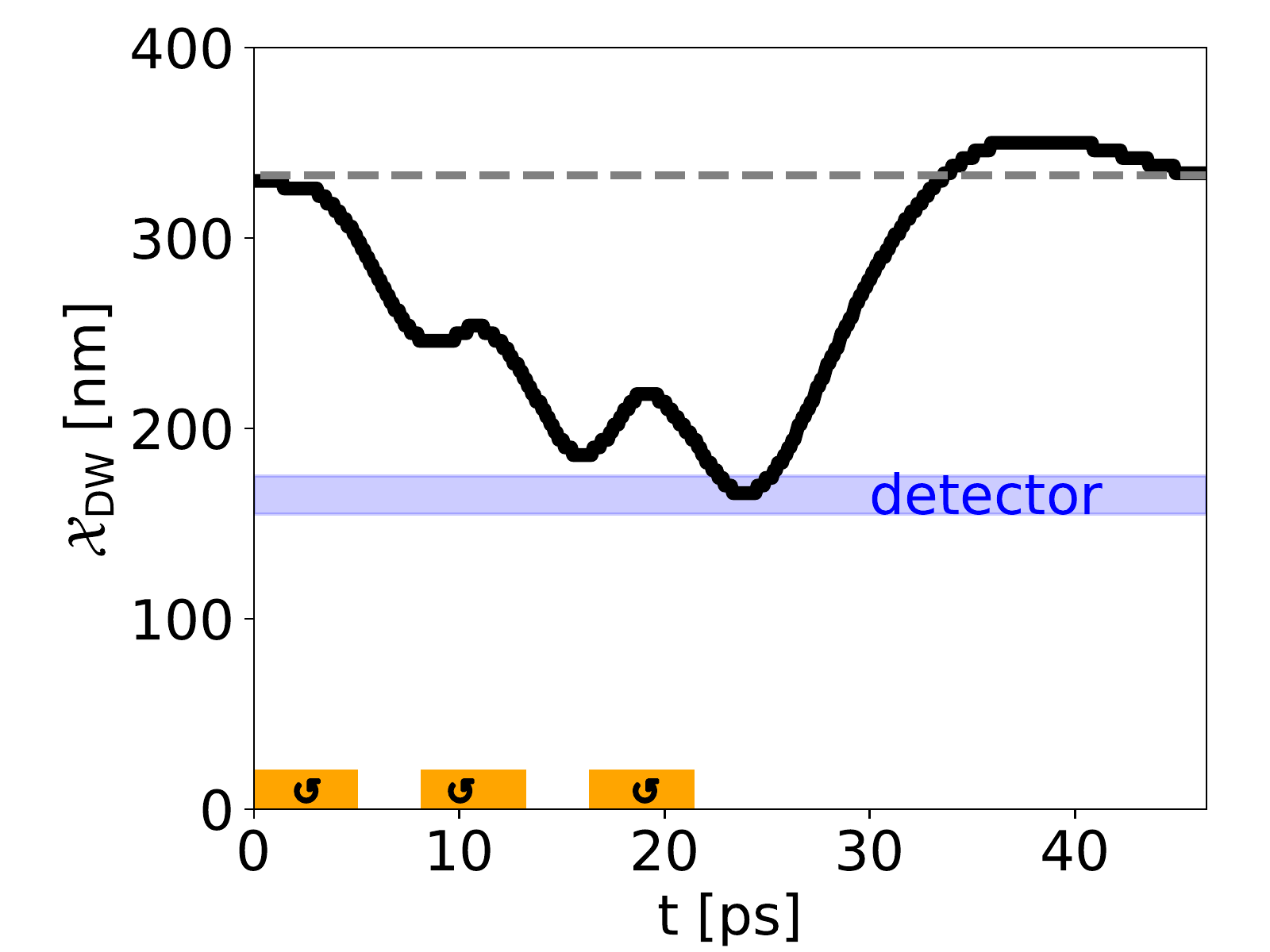}
    \put (13,60) {\small \textcolor{gray}{$\mathcal{X}_0$-}}
    \end{overpic}
    \subcaption{Trajectory of the AFM DW \label{fig:spikeDemotrajectoryIP}}
    \end{subfigure}\hfill%
    \begin{subfigure}{0.49\textwidth}%
    \centering%
    \includegraphics[width=1.0\linewidth]{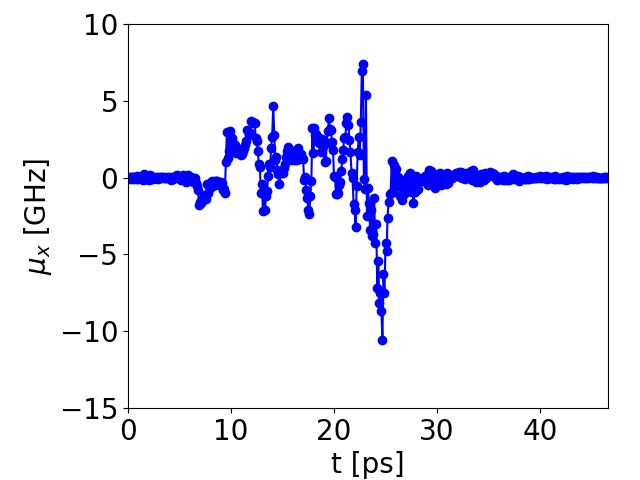}%
   \subcaption{Spin-pumping signal at the detector \label{fig:readoutSpikeIP}}%
    \end{subfigure}
    \caption{Leaky integrate-and-fire behavior of the all-magnonic AFM DW motion in the IP geometry with a DMI strength of $D=\SI{150}{\micro\joule\per\meter\squared}$. (a) The integration of three separate pulses, denoted by orange areas, provides enough energy to pull the DW away from its equilibrium position, denoted by the gray dashed line, to the detector, denoted by the blue area. This is the realization of the integrate-and-fire characteristic of the LIF model. During the inter-pulse intervals, the DW undergoes relaxation towards its equilibrium position, thereby exhibiting the \emph{leaky}  property. After the last pulse, the AFM DW relaxes back to the equilibrium position. (b) An impulse-like signal is fired when the DW passes the detector at t=\SI{25}{\pico\second}. This spike, generated when the synaptic inputs to the neuron reach a certain threshold value, represents the neuron action potential.}
    \label{fig:LeakyIntegrate}
\end{figure}

As we discussed earlier, biological neurons have LIF characteristics: if the input signal (or the sum of input spikes) reaches a threshold, the neuron fires, and then relaxes back to its ground state. In this part, we demonstrate that our proposed setup indeed can mimic the LIF behavior. In \cref{fig:spikeDemotrajectoryIP} the time-dependent AFM DW position in the IP geometry is shown, excited with three successive short magnetic field pulses.
One single pulse is not strong enough to move the AFM DW to the detector while three pulses can move the DW toward the detector, where it triggers a spike in the read-out (see \cref{fig:readoutSpikeIP}, more explanation in the next section). This is a demonstration of the {\it{integrative}}-and-{\it{fire}} behavior of our proposed non-volatile spintronic-based neuron. 
The \emph{leaky} nature of the neuron becomes evident as the DW reverts towards its equilibrium position, influenced by the magnetic anisotropy profile, in the absence of the stimulating signal.

\subsection{Electrical Readout of the AFM DW Position}
A detector on top of the AFM stripe measures the passing of the AFM DW by converting the spin-pumping signal, induced by AFM DW dynamics, to an electric voltage via either the inverse spin Hall effect \cite{SpinHallEffects} or recently discovered the inverse anomalous spin Hall effect \cite{doi:10.1021/acs.nanolett.8b02114}. 
In the former case, the detector is a nonmagnetic heavy metal and can only measure the component of spin-pumping signal parallel to the interface. In the latter case, the detector is a ferromagnetic metal with a strong spin-orbit coupling that can measure different components of the spin-pumping signal. 

The interfacial spin accumulation that arises from the DW-dynamics-induced spin-pumping, is given by \cite{ArneSpinPumping, PhysRevB.102.020408}, 
\begin{equation} \label{eq:spinAccum} \centering
    \bm{\mu}(t) := G_r^{\uparrow \downarrow} \big\langle\sum_{i= \text{A, B}}\big(\bm{m}_i(t,\bm{r})\times \dot{\bm{m}}_i(t,\bm{r})\big)\big\rangle,
\end{equation}
where $G_r^{\uparrow \downarrow}$ is the real part of the  spin mixing conductance \cite{ArneSpinMixingConductance} and $\langle ... \rangle$ denotes spatial average over the detector interface region.  In the present calculations, we have ignored the contribution of the imaginary part of the spin mixing conductance in the total spin accumulation. This latter is sensitive to the quality of interfaces and is negligible at disordered interfaces \cite{ArneSpinPumping}. In \cref{appendixImPart}, we demonstrate that the contribution of the imaginary part of the spin mixing conductance to the spin pumping signal in our setup is in general negligible.

In \cref{fig:readoutSpikeIP}, the temporal evolution of the spin accumulation signal $\mu_x(t)$ is presented for the IP geometry. In this example, as shown in \cref{fig:spikeDemotrajectoryIP} and described in the previous section, an AFM DW is pulled towards the detector with several small pulses. At the detector, the spin-pumping signal \cref{eq:spinAccum} is recorded over time. We subtract the background signal caused by the pumped magnons to find the filtered spin pumping signal arising from the AFM DW dynamics (blue curve). This signal clearly shows a maximum at around $t=\SI{15}{ps}$, which is when the AFM DW reaches the detector. 

\begin{figure}
    \centering
    \begin{overpic}[width=0.7\textwidth]{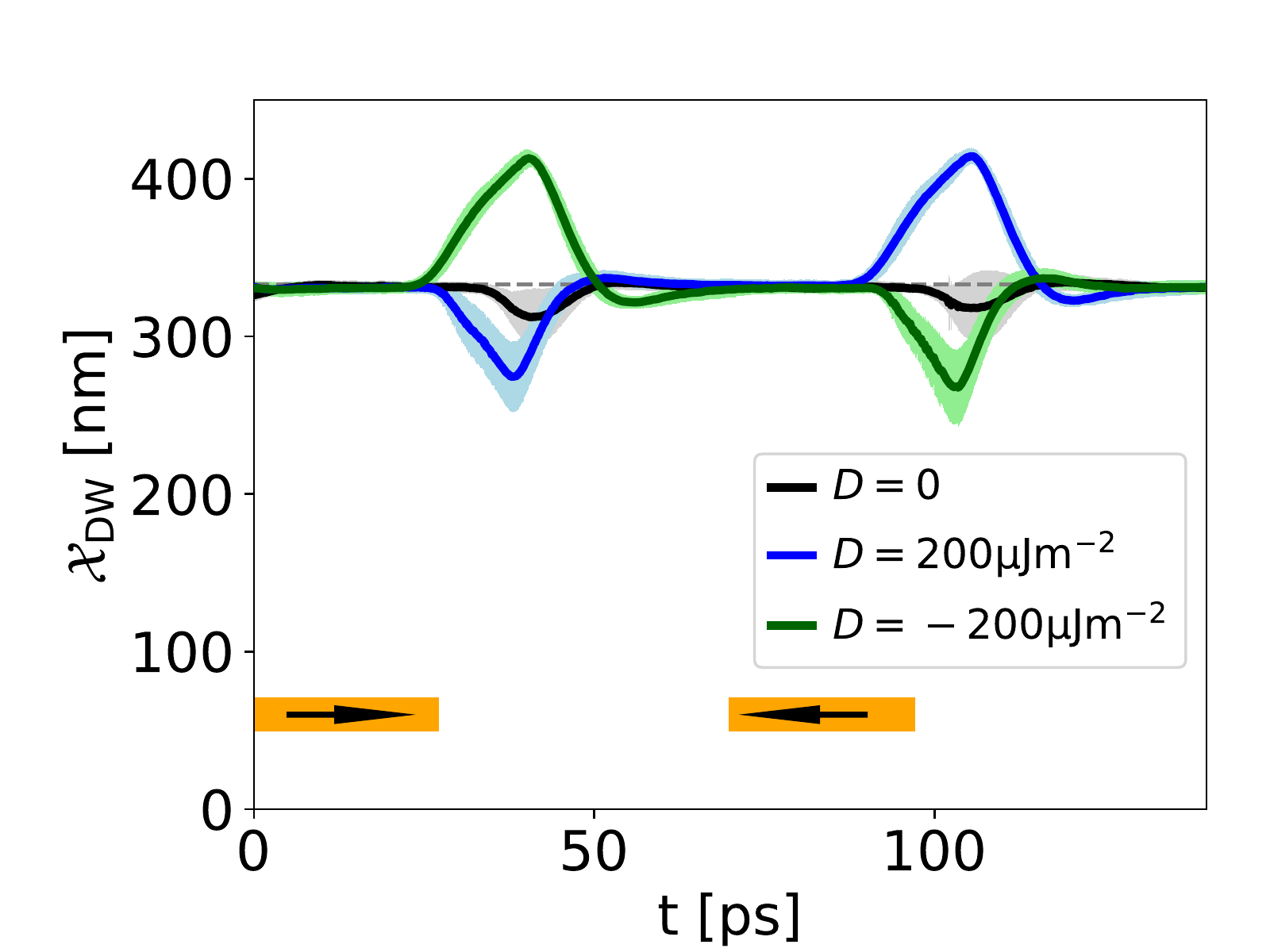}
    \put (15,51) {\textcolor{gray}{$\mathcal{X}_0$-}}
    \end{overpic}
    \caption{Electrical control of the AFM DW motion in the IP geometry. The orange areas depict the injector region that excites magnons via spin transfer torque pulses with two opposite spin torques, indicated by the arrow directions, at a finite temperature. Each trajectory is computed from an ensemble average over 60 realizations, and the uncertainty environment represents the standard deviation of the ensemble average. The equilibrium position of DW at $\mathcal{X}_0$ is denoted by a horizontal gray dashed line.}
    \label{fig:MoveDWwithTorque}
\end{figure}

\subsection{Magnon-Induced AFM-DW Motion by Spin Hall Torque} \label{sec:electricalControl}
Depending on the application, it might be an advantage to have an artificial single neuron that operates only electrically. To show our proposed setup has also all-electrical functionality, we replace the incident magnetic field pulse with a spin torque that results from a current-induced (anomalous) spin Hall torque in a non magnetic (magnetic) heavy-metal lead on top of the AFM at finite temperature. Through the (anomalous) spin Hall effect, a charge current in the injector is converted to a spin accumulation at the interface of the heavy metal and the AFM insulator. A nonequilibrium spin density with spin angular momentum along the easy-axis anisotropy may excite incoherent magnons in the AFM insulator via an interfacial spin transfer torque at finite temperature \cite{tunableLongDistanceKlaui,lebrun_long-distance_2020}. The chirality of excited magnons is controlled by the charge current direction and consequently the sign of the spin transfer torque.

Figure \ref{fig:MoveDWwithTorque} represents the displacement of an AFM DW in the IP geometry. Similar to the four-stage protocol used before, we run the following stages: After initialization of the DW in its equilibrium position $\mathcal{X}_0$, the spin transfer torque is turned on for \SI{25}{ps} as the first excitation stage, and then turned off for the first relaxation stage. 
In \cref{fig:MoveDWwithTorque} we see that the time interval between turning on the injector and the DW motion is much bigger compared to the previous case, where magnons were excited by a magnetic field, see \cref{fig:varyingDMI}. This is because the spin transfer torque excitation mechanism needs time to build up enough magnons in the system. 

In the second excitation stage, we change the sign of the spin accumulation and thus spin transfer torque in the injector, which is equivalent to reversing the direction of the charge current in the heavy metal layer. 
In \cref{fig:MoveDWwithTorque}, three AFM DW trajectories for different inhomogeneous DMI strengths are shown. Since temperature is finite and thus the time-evolution is non-deterministic, we perform an ensemble average for each AFM DW trajectory. The uncertainty environment for each line represents the standard deviation of the ensemble average. In the absence of DMI (black line), the direction of spin accumulation does not have an impact on the DW motion direction and the DW is pulled towards the injector in both cases. This is consistent with our previous result for magnon-induced by magnetic field case in which the direction of AFM DW motion was polarization independent in the absence of inhomogeneous DMI.
Turning the DMI on, however, leads to polarization-dependent DW motion. 

\subsection{Dynamical control of biologically realistic characteristics}
Recently, an artificial neuron based on AFM auto-oscillators was proposed \cite{Slavin2023Artificialneurons} and it was shown that this neuron owns some main ingredients of biological neurons.
In this subsection, we assess how our proposed neurons which are based on AFM DWs, intrinsically resemble some biological neurons characteristics, namely latency, bursting, inhibition, and refraction. We argue these features can be dynamically tuned  in our proposed model.

\textit{Neuronal response latency--} Latency describes the delay time between the excitation and the firing \cite{FundamentalNeuroscience}. In our proposed setup, this is the time between the excitation of magnons at the injector, and the read-out of the AFM-DW-induced spikes in the detector. This time is dependent on the excitation strength, the anisotropy profile, the distance of the detector and injector, and the material parameters. Thus, it can be tuned. In Figs. \ref{fig:varyingDMI}, \ref{fig:LeakyIntegrate} and \ref{fig:MoveDWwithTorque}, one can see the delay between the onset of the excitation (time window of excitation indicated by orange areas) and the DW movement.

\begin{figure}
    \centering \hfill
    \begin{subfigure}{0.49\textwidth}
    \centering
    \begin{overpic}[width=1\linewidth]{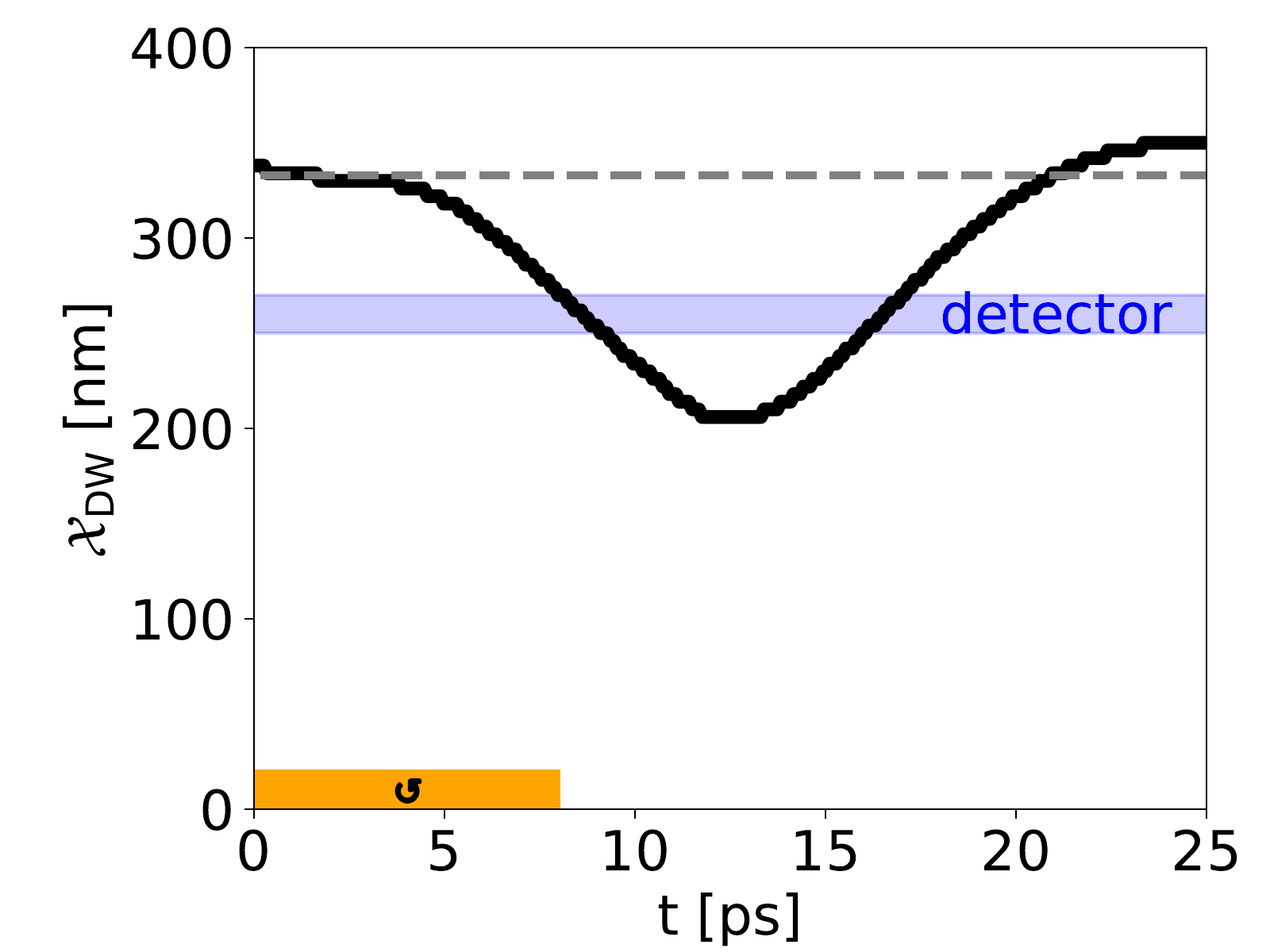}
    \put (12,60) {\textcolor{gray}{$\mathcal{X}_0$-}}
    \end{overpic}
    \subcaption{Trajectory of the AFM DW \label{fig:bursting1}}
    \end{subfigure}\hfill%
    \begin{subfigure}{0.49\textwidth}%
    \centering%
    \includegraphics[width=1\linewidth]{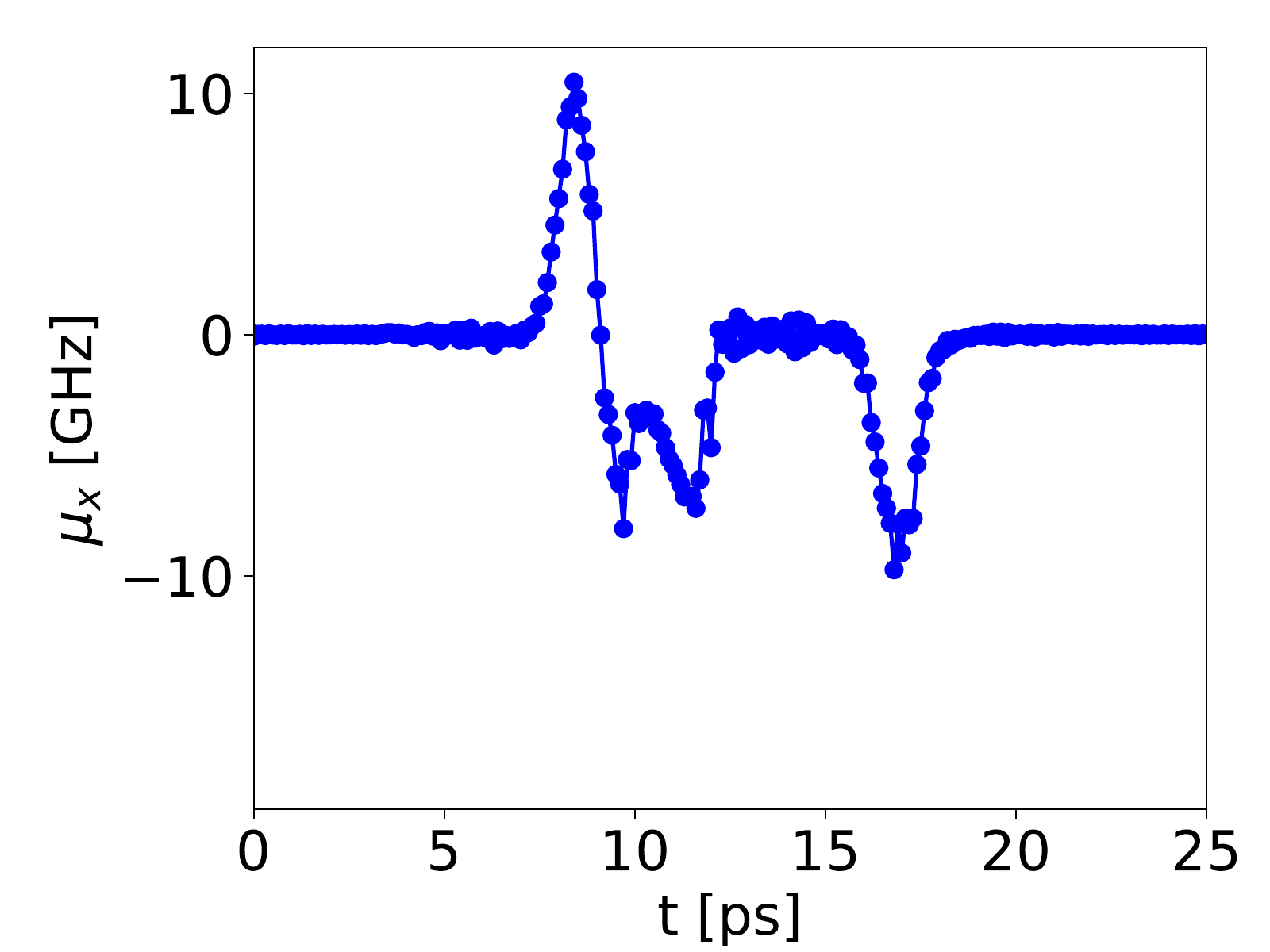}%
   \subcaption{Spin-pumping signal at the detector \label{fig:bursting2}}%
    \end{subfigure}
    \caption{Bursting behavior in the IP geometry with a DMI strength of $D=\SI{150}{\micro\joule\per\meter\squared}$. (a) A longer magnon excitation, here by a magnetic field, provides enough energy to pull the AFM-DW away from its equilibrium position, denoted by the gray dashed line, and passes the detector, denoted by the blue area. (b) An impulse-like signal with opposite polarity is fired each time the AFM-DW passes the detector in opposite directions. }
    \label{fig:bursting}
\end{figure}

\textit{Burst firing--} This is a dynamic state that happens when the input of neuron (or excitation strength) exceeds a certain threshold and, as a consequence, more than just one signal is fired \cite{Bursting}. In our system, this may happen when the DW is moved to greater distances from equilibrium compared to the detector distance. Then, it will pass underneath the detector twice, each time triggering an output signal. An example is shown in \cref{fig:bursting}, where the detector is placed closer to the equilibrium position compared to the case shown in \cref{sec:pump}. Like the latency, the bursting threshold is dependent on excitation strength, the magnetic anisotroy profile, detector distance and material parameters.
In \cref{fig:bursting2}, an additional signal present at around \SI{12}{ps} when DW passes the detector. We attribute this signal to the magnon emission by DW motion \cite{AFM-DWmagnonEmission,FM-DWmagnonEmission}.

\textit{Absolute refractory period--} The refractory period is the time that a neuron needs to relax back into the resting state from which it can fire again \cite{FundamentalNeuroscience}. In our system, the refractory period is non-zero if the DW passes the detector position (which happens in the case of bursting described before). Then, it has to relax back towards the equilibrium position before being able to fire again. 

\begin{figure}
    \centering%
    \begin{overpic}[width=0.5\linewidth]{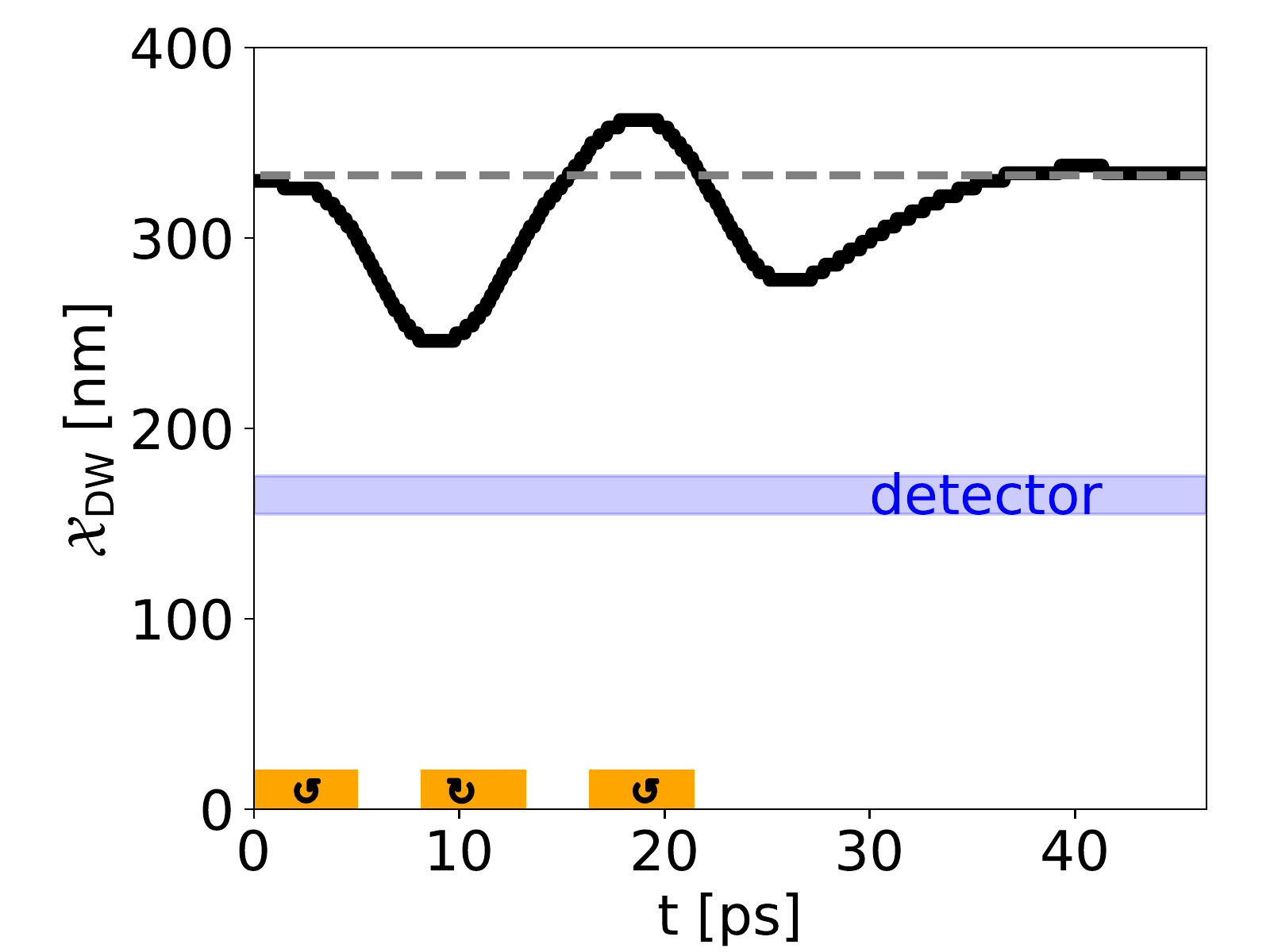}
    \put (12,60) {\textcolor{gray}{$\mathcal{X}_0$-}}
    \end{overpic}%
   \caption{Demonstration of inhibition in the IP geometry: Integration over excitation pulses with different helicities demonstrates the possibility of modelling inhibition. Compare to \cref{fig:spikeDemotrajectoryIP} where pulses with same chirality are integrated and lead to a spiking event.}\label{fig:DemoInhibition}%
    \end{figure}

\textit{Neural inhibition--} Biological neurons can exert inhibitory control over their connected neurons. Inhibitory neurons modulate the firing behavior of other neurons, signaling them to refrain from firing \cite{Inhibition}.
In the network structure, inhibition corresponds to negative weights \cite{inhibitionNegWeights}. In our proposal, negative weights can be achieved by placing a detector to the left and right of the equilibrium position of the DW.
As demonstrated in  \cref{fig:VaryingD-IP} and \cref{fig:MoveDWwithTorque}, the helicity of the applied magnetic field and the direction of the spin torque control the direction of the DW displacement, determine whether the signal is detected at the left or right detector during spike readout. Subsequently, it becomes feasible to attribute a positive weight to one of the readout signals and a negative weight to the other. Consequently, upon integration into the subsequent layer, these weights correspond to the helicity or spin torque direction. Thus, during the integration of pulses in the next neuron, competing forces can act on the DW.
An example is shown in \cref{fig:DemoInhibition} where one of the tree excitation pulses has opposite chirality and thus pushes the DW away from the detector. To the best of  knowledge, inhibition has not been incorporated in FM DW-based neurons thus far. However, as demonstrated in our proposal, the chirality of magnons in AFM systems represents a crucial degree of freedom that enables this particular feature of biological neurons.

\subsection{Suggested network structure}
In this article, a detailed study of an AFM-based single neuron was conducted, focusing on the demonstration of its LIF properties. Although further implementations extend beyond the scope of this work, a brief outlook will be provided on the construction of a SNN using the proposed neurons.\\
As explained in \cref{sec:snn}, the input to each neuron involves the accumulation of multiple spike trains. In our system, this process is modeled using pulses of either a magnetic field or an electric current-induced SHE. Within the neuron, the integration of incoming pulses may or may not lead to a spiking event. The output is an electrical readout of the spiking event, which is subsequently forwarded to the next layer.
To facilitate network training, incoming signals can be scaled with trainable weights, denoted as $w_i$, see \cref{fig:SNN}. In our system, the amplitude of these weights corresponds to the excitation strength and/or duration, while the sign can be set by evaluating which detector reads out the spike. If the weight is negative, in the subsequent neuron, magnons of opposite chirality are excited by reversing the helicity of the magnetic field or the current direction respectively.

\section*{Summary and Concluding Remarks}\label{section:discussion}
In this paper, we have proposed a non-volatile, low-energy cost, and fast operating single neuron, which is based on a DW texture in an AFM insulator with an anisotropy gradient. Our proposed AFM-based neuron shows a leaky integrated-fire behavior, which can model a biological neuron. This single neuron is activated by AFM magnons, which can be excited at the source region by either a magnetic field pulse or spin transfer torque mechanism. 
The source region that injects magnons into the system resembles a dendrite in a nerve cell.

Our proposed AFM-based single neuron can have two detectors, which makes it possible to model inhibition feature of biological neurons. The detectors act as transmitters, resembling synaptic terminals of neurons, and will be connected to neighboring neurons.
In general, one can replace the AFM DW in our setup with topologically stable AFM skyrmions as well. 
Synchronization and functionality of the connected single neurons remain as an important open question that should be explored further theoretically and experimentally in next studies.


\bibliography{literature}

\begin{thebibliography}{10}
\urlstyle{rm}
\expandafter\ifx\csname url\endcsname\relax
  \def\url#1{\texttt{#1}}\fi
\expandafter\ifx\csname urlprefix\endcsname\relax\def\urlprefix{URL }\fi
\expandafter\ifx\csname doiprefix\endcsname\relax\def\doiprefix{DOI: }\fi
\providecommand{\bibinfo}[2]{#2}
\providecommand{\eprint}[2][]{\url{#2}}

\bibitem{mahmoud_introduction_2020}
\bibinfo{author}{Mahmoud, A.} \emph{et~al.}
\newblock \bibinfo{journal}{\bibinfo{title}{An introduction to spin wave
  computing}}.
\newblock {\emph{\JournalTitle{J. Appl. Phys.}}}
  \textbf{\bibinfo{volume}{128}}, \bibinfo{pages}{161101},
  \doiprefix\url{10.1063/5.0019328} (\bibinfo{year}{2020}).

\bibitem{Christensen_2022}
\bibinfo{author}{Christensen, D.~V.} \emph{et~al.}
\newblock \bibinfo{journal}{\bibinfo{title}{2022 roadmap on neuromorphic
  computing and engineering}}.
\newblock {\emph{\JournalTitle{Neuromorph. Comput. Eng.}}}
  \textbf{\bibinfo{volume}{2}}, \bibinfo{pages}{022501},
  \doiprefix\url{10.1088/2634-4386/ac4a83} (\bibinfo{year}{2022}).

\bibitem{Low-Power-Computing}
\bibinfo{author}{Liu, D.}, \bibinfo{author}{Yu, H.} \& \bibinfo{author}{Chai,
  Y.}
\newblock \bibinfo{journal}{\bibinfo{title}{Low-power computing with
  neuromorphic engineering}}.
\newblock {\emph{\JournalTitle{Adv. Intell. Syst.}}}
  \textbf{\bibinfo{volume}{3}}, \bibinfo{pages}{2000150},
  \doiprefix\url{10.1002/aisy.202000150} (\bibinfo{year}{2020}).

\bibitem{Spike-based-Neuromorphic}
\bibinfo{author}{Rao, A.}, \bibinfo{author}{Plank, P.}, \bibinfo{author}{Wild,
  A.} \& \bibinfo{author}{Maass, W.}
\newblock \bibinfo{journal}{\bibinfo{title}{A long short-term memory for ai
  applications in spike-based neuromorphic hardware}}.
\newblock {\emph{\JournalTitle{Nat. Mach. Intell.}}}
  \textbf{\bibinfo{volume}{4}}, \bibinfo{pages}{467},
  \doiprefix\url{10.1038/s42256-022-00480-w} (\bibinfo{year}{2022}).

\bibitem{Communication-consumes}
\bibinfo{author}{Levya, W.~B.} \& \bibinfo{author}{Calvert, V.~G.}
\newblock \bibinfo{journal}{\bibinfo{title}{Communication consumes 35 times
  more energy than computation in the human cortex, but both costs are needed
  to predict synapse number}}.
\newblock {\emph{\JournalTitle{Proc. Natl. Acad. Sci. U.S.A.}}}
  \textbf{\bibinfo{volume}{118}}, \bibinfo{pages}{e2008173118},
  \doiprefix\url{10.1073/pnas.2008173118} (\bibinfo{year}{2021}).

\bibitem{roadmap}
\bibinfo{author}{An, H.}, \bibinfo{author}{Bai, K.} \& \bibinfo{author}{Yi, Y.}
\newblock \emph{\bibinfo{title}{The Roadmap to Realizing Memristive
  Three-Dimensional Neuromorphic Computing System Advances in Memristor Neural
  Networks - Modeling and Applications}} (\bibinfo{publisher}{IntechOpen},
  \bibinfo{year}{2018}).

\bibitem{grollier_neuromorphic_2020}
\bibinfo{author}{Grollier, J.} \emph{et~al.}
\newblock \bibinfo{journal}{\bibinfo{title}{Neuromorphic spintronics}}.
\newblock {\emph{\JournalTitle{Nat. Electron.}}} \textbf{\bibinfo{volume}{3}},
  \bibinfo{pages}{360}, \doiprefix\url{10.1038/s41928-019-0360-9}
  (\bibinfo{year}{2020}).

\bibitem{BRATAAS20201}
\bibinfo{author}{Brataas, A.}, \bibinfo{author}{{van Wees}, B.},
  \bibinfo{author}{Klein, O.}, \bibinfo{author}{{de Loubens}, G.} \&
  \bibinfo{author}{Viret, M.}
\newblock \bibinfo{journal}{\bibinfo{title}{Spin insulatronics}}.
\newblock {\emph{\JournalTitle{Phys. Rep.}}} \textbf{\bibinfo{volume}{885}},
  \bibinfo{pages}{1--27},
  \doiprefix\url{https://doi.org/10.1016/j.physrep.2020.08.006}
  (\bibinfo{year}{2020}).
\newblock \bibinfo{note}{Spin Insulatronics}.

\bibitem{IEEEreviewFM}
\bibinfo{author}{Brigner, W.~H.} \emph{et~al.}
\newblock \bibinfo{title}{Purely spintronic leaky integrate-and-fire neurons}.
\newblock In \emph{\bibinfo{booktitle}{2022 IEEE International Symposium on
  Circuits and Systems (ISCAS)}}, \bibinfo{pages}{1189--1193},
  \doiprefix\url{10.1109/ISCAS48785.2022.9937890} (\bibinfo{year}{2022}).

\bibitem{IEEEbringerMTJ}
\bibinfo{author}{Brigner, W.~H.} \emph{et~al.}
\newblock \bibinfo{journal}{\bibinfo{title}{Domain wall leaky
  integrate-and-fire neurons with shape-based configurable activation
  functions}}.
\newblock {\emph{\JournalTitle{IEEE Transactions on Electron Devices}}}
  \textbf{\bibinfo{volume}{69}}, \bibinfo{pages}{2353--2359},
  \doiprefix\url{10.1109/TED.2022.3159508} (\bibinfo{year}{2022}).

\bibitem{BringerAnisoGradient}
\bibinfo{author}{Brigner, W.~H.} \emph{et~al.}
\newblock \bibinfo{journal}{\bibinfo{title}{Graded-anisotropy-induced magnetic
  domain wall drift for an artificial spintronic leaky integrate-and-fire
  neuron}}.
\newblock {\emph{\JournalTitle{IEEE Journal on Exploratory Solid-State
  Computational Devices and Circuits}}} \textbf{\bibinfo{volume}{5}},
  \bibinfo{pages}{19--24}, \doiprefix\url{10.1109/JXCDC.2019.2904191}
  (\bibinfo{year}{2019}).

\bibitem{Antiferromagnetic_spintronics}
\bibinfo{author}{Jungwirth, T.}, \bibinfo{author}{Marti, X.},
  \bibinfo{author}{Wadley, P.} \& \bibinfo{author}{Wunderlich}.
\newblock \bibinfo{journal}{\bibinfo{title}{Antiferromagnetic spintronics}}.
\newblock {\emph{\JournalTitle{Nat. Nanotechnol.}}}
  \textbf{\bibinfo{volume}{11}}, \bibinfo{pages}{231},
  \doiprefix\url{10.1038/nnano.2016.18} (\bibinfo{year}{2016}).

\bibitem{Neuromorphic_computing}
\bibinfo{author}{Kurenkov, A.}, \bibinfo{author}{Fukami, S.} \&
  \bibinfo{author}{Ohno, H.}
\newblock \bibinfo{journal}{\bibinfo{title}{Neuromorphic computing with
  antiferromagnetic spintronics}}.
\newblock {\emph{\JournalTitle{J. Appl. Phys.}}}
  \textbf{\bibinfo{volume}{128}}, \bibinfo{pages}{010902},
  \doiprefix\url{10.1063/5.0009482} (\bibinfo{year}{2020}).

\bibitem{PhysRevLett.125.207202}
\bibinfo{author}{Zhang, S.} \& \bibinfo{author}{Tserkovnyak, Y.}
\newblock \bibinfo{journal}{\bibinfo{title}{Antiferromagnet-based neuromorphics
  using dynamics of topological charges}}.
\newblock {\emph{\JournalTitle{Phys. Rev. Lett.}}}
  \textbf{\bibinfo{volume}{125}}, \bibinfo{pages}{207202},
  \doiprefix\url{10.1103/PhysRevLett.125.207202} (\bibinfo{year}{2020}).

\bibitem{Artificial_neurons}
\bibinfo{author}{Bradley, H.} \emph{et~al.}
\newblock \bibinfo{journal}{\bibinfo{title}{Artificial neurons based on
  antiferromagnetic auto-oscillators as a platform for neuromorphic
  computing}}.
\newblock {\emph{\JournalTitle{arXiv}}}
  \doiprefix\url{10.48550/ARXIV.2208.06565} (\bibinfo{year}{2022}).

\bibitem{Bindal_2021}
\bibinfo{author}{Bindal, N.}, \bibinfo{author}{Ian, C. A.~C.},
  \bibinfo{author}{Lew, W.~S.} \& \bibinfo{author}{Kaushik, B.~K.}
\newblock \bibinfo{journal}{\bibinfo{title}{Antiferromagnetic skyrmion
  repulsion based artificial neuron device}}.
\newblock {\emph{\JournalTitle{Nanotechnology}}} \textbf{\bibinfo{volume}{32}},
  \bibinfo{pages}{215204}, \doiprefix\url{10.1088/1361-6528/abe261}
  (\bibinfo{year}{2021}).

\bibitem{MAASS19971659}
\bibinfo{author}{Maass, W.}
\newblock \bibinfo{journal}{\bibinfo{title}{Networks of spiking neurons: The
  third generation of neural network models}}.
\newblock {\emph{\JournalTitle{Neural Netw.}}} \textbf{\bibinfo{volume}{10}},
  \bibinfo{pages}{1659--1671},
  \doiprefix\url{https://doi.org/10.1016/S0893-6080(97)00011-7}
  (\bibinfo{year}{1997}).

\bibitem{gerstner_spiking_2002}
\bibinfo{author}{Gerstner, W.} \& \bibinfo{author}{Kistler, W.~M.}
\newblock \emph{\bibinfo{title}{Spiking Neuron Models: Single Neurons,
  Populations, Plasticity}} (\bibinfo{publisher}{Cambridge University Press},
  \bibinfo{year}{2002}).

\bibitem{STEIN1965173}
\bibinfo{author}{Stein, R.~B.}
\newblock \bibinfo{journal}{\bibinfo{title}{A theoretical analysis of neuronal
  variability}}.
\newblock {\emph{\JournalTitle{Biophys. J.}}} \textbf{\bibinfo{volume}{5}},
  \bibinfo{pages}{173--194},
  \doiprefix\url{https://doi.org/10.1016/S0006-3495(65)86709-1}
  (\bibinfo{year}{1965}).

\bibitem{https://doi.org/10.1002/adfm.201604740}
\bibinfo{author}{Stoliar, P.} \emph{et~al.}
\newblock \bibinfo{journal}{\bibinfo{title}{A leaky-integrate-and-fire neuron
  analog realized with a mott insulator}}.
\newblock {\emph{\JournalTitle{Adv. Funct. Mater.}}}
  \textbf{\bibinfo{volume}{27}}, \bibinfo{pages}{1604740},
  \doiprefix\url{https://doi.org/10.1002/adfm.201604740}
  (\bibinfo{year}{2017}).

\bibitem{maass_networks_1997}
\bibinfo{author}{Maass, W.}
\newblock \bibinfo{journal}{\bibinfo{title}{Networks of spiking neurons: The
  third generation of neural network models}}.
\newblock {\emph{\JournalTitle{Neural Netw.}}} \textbf{\bibinfo{volume}{10}},
  \bibinfo{pages}{1659}, \doiprefix\url{10.1016/S0893-6080(97)00011-7}
  (\bibinfo{year}{1997}).

\bibitem{jang_spiking_2021}
\bibinfo{author}{Jang, H.}, \bibinfo{author}{Skatchkovsky, N.} \&
  \bibinfo{author}{Simeone, O.}
\newblock \bibinfo{journal}{\bibinfo{title}{Spiking neural networks-part i:
  Detecting spatial patterns}}.
\newblock {\emph{\JournalTitle{IEEE Commun. Lett.}}}
  \textbf{\bibinfo{volume}{25}}, \bibinfo{pages}{1736},
  \doiprefix\url{10.1109/LCOMM.2021.3050207} (\bibinfo{year}{2021}).

\bibitem{bohte_error-backpropagation_2002}
\bibinfo{author}{Bohte, S.~M.}, \bibinfo{author}{Kok, J.~N.} \&
  \bibinfo{author}{La~Poutré, H.}
\newblock \bibinfo{journal}{\bibinfo{title}{Error-backpropagation in temporally
  encoded networks of spiking neurons}}.
\newblock {\emph{\JournalTitle{Neurocomputing (Amsterdam)}}}
  \textbf{\bibinfo{volume}{48}}, \bibinfo{pages}{17},
  \doiprefix\url{10.1016/S0925-2312(01)00658-0} (\bibinfo{year}{2002}).

\bibitem{AtomMicroScales}
\bibinfo{author}{Etz, C.}, \bibinfo{author}{Bergqvist, L.},
  \bibinfo{author}{Bergman, A.}, \bibinfo{author}{Taroni, A.} \&
  \bibinfo{author}{Eriksson, O.}
\newblock \bibinfo{journal}{\bibinfo{title}{Atomistic spin dynamics and surface
  magnons}}.
\newblock {\emph{\JournalTitle{J. Phys. Condens. Matter}}}
  \textbf{\bibinfo{volume}{27}}, \bibinfo{pages}{243202},
  \doiprefix\url{10.1088/0953-8984/27/24/243202} (\bibinfo{year}{2015}).

\bibitem{lepadatu_boris_2020}
\bibinfo{author}{Lepadatu, S.}
\newblock \bibinfo{journal}{\bibinfo{title}{{Boris computational
  spintronics—High performance multi-mesh magnetic and spin transport
  modeling software}}}.
\newblock {\emph{\JournalTitle{J. Appl. Phys.}}}
  \textbf{\bibinfo{volume}{128}}, \bibinfo{pages}{243902},
  \doiprefix\url{10.1063/5.0024382} (\bibinfo{year}{2020}).

\bibitem{PhysRevLett.110.127208}
\bibinfo{author}{Tveten, E.~G.}, \bibinfo{author}{Qaiumzadeh, A.},
  \bibinfo{author}{Tretiakov, O.~A.} \& \bibinfo{author}{Brataas, A.}
\newblock \bibinfo{journal}{\bibinfo{title}{Staggered dynamics in
  antiferromagnets by collective coordinates}}.
\newblock {\emph{\JournalTitle{Phys. Rev. Lett.}}}
  \textbf{\bibinfo{volume}{110}}, \bibinfo{pages}{127208},
  \doiprefix\url{10.1103/PhysRevLett.110.127208} (\bibinfo{year}{2013}).

\bibitem{rana_towards_2019}
\bibinfo{author}{Rana, B.} \& \bibinfo{author}{Otani, Y.}
\newblock \bibinfo{journal}{\bibinfo{title}{Towards magnonic devices based on
  voltage-controlled magnetic anisotropy}}.
\newblock {\emph{\JournalTitle{Commun. Phys.}}} \textbf{\bibinfo{volume}{2}},
  \bibinfo{pages}{90}, \doiprefix\url{10.1038/s42005-019-0189-6}
  (\bibinfo{year}{2019}).

\bibitem{VCMA1}
\bibinfo{author}{Ma, C.} \emph{et~al.}
\newblock \bibinfo{journal}{\bibinfo{title}{Electric field-induced creation and
  directional motion of domain walls and skyrmion bubbles}}.
\newblock {\emph{\JournalTitle{Nano Lett.}}} \textbf{\bibinfo{volume}{19}},
  \bibinfo{pages}{353}, \doiprefix\url{10.1021/acs.nanolett.8b03983}
  (\bibinfo{year}{2018}).

\bibitem{VCMA2}
\bibinfo{author}{Yu, G.} \emph{et~al.}
\newblock \bibinfo{journal}{\bibinfo{title}{Room-temperature creation and
  spin–orbit torque manipulation of skyrmions in thin films with engineered
  asymmetry}}.
\newblock {\emph{\JournalTitle{Nano Lett.}}} \textbf{\bibinfo{volume}{16}},
  \bibinfo{pages}{1981}, \doiprefix\url{10.1021/acs.nanolett.5b05257}
  (\bibinfo{year}{2016}).

\bibitem{VCMA3}
\bibinfo{author}{Skowroński, W.} \emph{et~al.}
\newblock \bibinfo{journal}{\bibinfo{title}{Perpendicular magnetic anisotropy
  of ir/cofeb/mgo trilayer system tuned by electric fields}}.
\newblock {\emph{\JournalTitle{Appl. Phys. Express}}}
  \textbf{\bibinfo{volume}{8}}, \bibinfo{pages}{053003},
  \doiprefix\url{10.7567/APEX.8.053003} (\bibinfo{year}{2015}).

\bibitem{VCMA4}
\bibinfo{author}{Kawabe, T.} \emph{et~al.}
\newblock \bibinfo{journal}{\bibinfo{title}{Electric-field-induced changes of
  magnetic moments and magnetocrystalline anisotropy in ultrathin cobalt
  films}}.
\newblock {\emph{\JournalTitle{Phys. Rev. B}}} \textbf{\bibinfo{volume}{96}},
  \bibinfo{pages}{220412}, \doiprefix\url{10.1103/PhysRevB.96.220412}
  (\bibinfo{year}{2017}).

\bibitem{https://doi.org/10.1002/advs.201800356}
\bibinfo{author}{Herklotz, A.} \emph{et~al.}
\newblock \bibinfo{journal}{\bibinfo{title}{Designing magnetic anisotropy
  through strain doping}}.
\newblock {\emph{\JournalTitle{Adv. Sci.}}} \textbf{\bibinfo{volume}{5}},
  \bibinfo{pages}{1800356},
  \doiprefix\url{https://doi.org/10.1002/advs.201800356}
  (\bibinfo{year}{2018}).

\bibitem{https://doi.org/10.1002/pssr.201900467}
\bibinfo{author}{Belyaev, B.~A.}, \bibinfo{author}{Izotov, A.~V.},
  \bibinfo{author}{Solovev, P.~N.} \& \bibinfo{author}{Boev, N.~M.}
\newblock \bibinfo{journal}{\bibinfo{title}{Strain-gradient-induced
  unidirectional magnetic anisotropy in nanocrystalline thin permalloy films}}.
\newblock {\emph{\JournalTitle{Phys. Status Solidi RRL}}}
  \textbf{\bibinfo{volume}{14}}, \bibinfo{pages}{1900467},
  \doiprefix\url{https://doi.org/10.1002/pssr.201900467}
  (\bibinfo{year}{2020}).

\bibitem{https://doi.org/10.48550/arxiv.2209.04527}
\bibinfo{author}{Ebrahimian, A.}, \bibinfo{author}{Dyrdał, A.} \&
  \bibinfo{author}{Qaiumzadeh, A.}
\newblock \bibinfo{journal}{\bibinfo{title}{Control of magnetic states and spin
  interactions in bilayer crcl$_{3}$ with strain and electric fields}}.
\newblock {\emph{\JournalTitle{arXiv}}}
  \doiprefix\url{10.48550/ARXIV.2209.04527} (\bibinfo{year}{2022}).

\bibitem{PhysRevMaterials.4.094004}
\bibinfo{author}{Vishkayi, S.~I.}, \bibinfo{author}{Torbatian, Z.},
  \bibinfo{author}{Qaiumzadeh, A.} \& \bibinfo{author}{Asgari, R.}
\newblock \bibinfo{journal}{\bibinfo{title}{Strain and electric-field control
  of spin-spin interactions in monolayer ${\mathrm{cri}}_{3}$}}.
\newblock {\emph{\JournalTitle{Phys. Rev. Materials}}}
  \textbf{\bibinfo{volume}{4}}, \bibinfo{pages}{094004},
  \doiprefix\url{10.1103/PhysRevMaterials.4.094004} (\bibinfo{year}{2020}).

\bibitem{Meer_StrainInducedAniso}
\bibinfo{author}{Meer, H.} \emph{et~al.}
\newblock \bibinfo{journal}{\bibinfo{title}{Strain-induced shape anisotropy in
  antiferromagnetic structures}}.
\newblock {\emph{\JournalTitle{Phys. Rev. B}}} \textbf{\bibinfo{volume}{106}},
  \bibinfo{pages}{094430}, \doiprefix\url{10.1103/PhysRevB.106.094430}
  (\bibinfo{year}{2022}).

\bibitem{AnisoProfileStress}
\bibinfo{author}{Zhukova, V.} \emph{et~al.}
\newblock \bibinfo{journal}{\bibinfo{title}{Grading the magnetic anisotropy and
  engineering the domain wall dynamics in fe-rich microwires by
  stress-annealing}}.
\newblock {\emph{\JournalTitle{Acta Materialia}}}
  \textbf{\bibinfo{volume}{155}}, \bibinfo{pages}{279--285},
  \doiprefix\url{https://doi.org/10.1016/j.actamat.2018.05.068}
  (\bibinfo{year}{2018}).

\bibitem{tunableLongDistanceKlaui}
\bibinfo{author}{Lebrun, R.} \emph{et~al.}
\newblock \bibinfo{journal}{\bibinfo{title}{Tunable long-distance spin
  transport in a crystalline antiferromagnetic iron oxide}}.
\newblock {\emph{\JournalTitle{Nature}}} \textbf{\bibinfo{volume}{561}},
  \bibinfo{pages}{222}, \doiprefix\url{10.1038/s41586-018-0490-7}
  (\bibinfo{year}{2018}).

\bibitem{doi:10.1021/acs.nanolett.8b02114}
\bibinfo{author}{Das, K.~S.}, \bibinfo{author}{Liu, J.}, \bibinfo{author}{van
  Wees, B.~J.} \& \bibinfo{author}{Vera-Marun, I.~J.}
\newblock \bibinfo{journal}{\bibinfo{title}{Efficient injection and detection
  of out-of-plane spins via the anomalous spin hall effect in permalloy
  nanowires}}.
\newblock {\emph{\JournalTitle{Nano Lett.}}} \textbf{\bibinfo{volume}{18}},
  \bibinfo{pages}{5633}, \doiprefix\url{10.1021/acs.nanolett.8b02114}
  (\bibinfo{year}{2018}).

\bibitem{ArneSpinPumping}
\bibinfo{author}{Cheng, R.}, \bibinfo{author}{Xiao, J.}, \bibinfo{author}{Niu,
  Q.} \& \bibinfo{author}{Brataas, A.}
\newblock \bibinfo{journal}{\bibinfo{title}{Spin pumping and spin-transfer
  torques in antiferromagnets}}.
\newblock {\emph{\JournalTitle{Phys. Rev. Lett.}}}
  \textbf{\bibinfo{volume}{113}}, \bibinfo{pages}{057601},
  \doiprefix\url{10.1103/PhysRevLett.113.057601} (\bibinfo{year}{2014}).

\bibitem{https://doi.org/10.48550/arxiv.2211.01195}
\bibinfo{author}{Mal'shukov, A.~G.}
\newblock \bibinfo{journal}{\bibinfo{title}{Spin pumping by a moving domain
  wall at the interface of an antiferromagnetic insulator and a two-dimensional
  metal}}.
\newblock {\emph{\JournalTitle{arXiv}}}
  \doiprefix\url{10.48550/ARXIV.2211.01195} (\bibinfo{year}{2022}).

\bibitem{doi:10.1063/1.4967171}
\bibinfo{author}{Pham, V.~T.} \emph{et~al.}
\newblock \bibinfo{journal}{\bibinfo{title}{Electrical detection of magnetic
  domain walls by inverse and direct spin hall effect}}.
\newblock {\emph{\JournalTitle{Appl. Phys. Lett.}}}
  \textbf{\bibinfo{volume}{109}}, \bibinfo{pages}{192401},
  \doiprefix\url{10.1063/1.4967171} (\bibinfo{year}{2016}).

\bibitem{AlirezaHelicityAFMdw}
\bibinfo{author}{Qaiumzadeh, A.}, \bibinfo{author}{Kristiansen, L.~A.} \&
  \bibinfo{author}{Brataas, A.}
\newblock \bibinfo{journal}{\bibinfo{title}{Controlling chiral domain walls in
  antiferromagnets using spin-wave helicity}}.
\newblock {\emph{\JournalTitle{Phys. Rev. B}}} \textbf{\bibinfo{volume}{97}},
  \bibinfo{pages}{020402}, \doiprefix\url{10.1103/PhysRevB.97.020402}
  (\bibinfo{year}{2018}).

\bibitem{agrawal_mimicking_2019}
\bibinfo{author}{Agrawal, A.} \& \bibinfo{author}{Roy, K.}
\newblock \bibinfo{journal}{\bibinfo{title}{Mimicking leaky-integrate-fire
  spiking neuron using automotion of domain walls for energy-efficient
  brain-inspired computing}}.
\newblock {\emph{\JournalTitle{IEEE Trans. Magn.}}}
  \textbf{\bibinfo{volume}{55}}, \bibinfo{pages}{1},
  \doiprefix\url{10.1109/TMAG.2018.2882164} (\bibinfo{year}{2019}).

\bibitem{ArneSpinMixingConductance}
\bibinfo{author}{Tserkovnyak, Y.}, \bibinfo{author}{Brataas, A.} \&
  \bibinfo{author}{Bauer, G. E.~W.}
\newblock \bibinfo{journal}{\bibinfo{title}{Enhanced gilbert damping in thin
  ferromagnetic films}}.
\newblock {\emph{\JournalTitle{Phys. Rev. Lett.}}}
  \textbf{\bibinfo{volume}{88}}, \bibinfo{pages}{117601},
  \doiprefix\url{10.1103/PhysRevLett.88.117601} (\bibinfo{year}{2002}).

\bibitem{PhysRevLett.111.097602}
\bibinfo{author}{Kapelrud, A.} \& \bibinfo{author}{Brataas, A.}
\newblock \bibinfo{journal}{\bibinfo{title}{Spin pumping and enhanced gilbert
  damping in thin magnetic insulator films}}.
\newblock {\emph{\JournalTitle{Phys. Rev. Lett.}}}
  \textbf{\bibinfo{volume}{111}}, \bibinfo{pages}{097602},
  \doiprefix\url{10.1103/PhysRevLett.111.097602} (\bibinfo{year}{2013}).

\bibitem{PhysRevLett.112.147204}
\bibinfo{author}{Tveten, E.~G.}, \bibinfo{author}{Qaiumzadeh, A.} \&
  \bibinfo{author}{Brataas, A.}
\newblock \bibinfo{journal}{\bibinfo{title}{Antiferromagnetic domain wall
  motion induced by spin waves}}.
\newblock {\emph{\JournalTitle{Phys. Rev. Lett.}}}
  \textbf{\bibinfo{volume}{112}}, \bibinfo{pages}{147204},
  \doiprefix\url{10.1103/PhysRevLett.112.147204} (\bibinfo{year}{2014}).

\bibitem{PhysRevB.99.054423}
\bibinfo{author}{Khoshlahni, R.}, \bibinfo{author}{Qaiumzadeh, A.},
  \bibinfo{author}{Bergman, A.} \& \bibinfo{author}{Brataas, A.}
\newblock \bibinfo{journal}{\bibinfo{title}{Ultrafast generation and dynamics
  of isolated skyrmions in antiferromagnetic insulators}}.
\newblock {\emph{\JournalTitle{Phys. Rev. B}}} \textbf{\bibinfo{volume}{99}},
  \bibinfo{pages}{054423}, \doiprefix\url{10.1103/PhysRevB.99.054423}
  (\bibinfo{year}{2019}).

\bibitem{AFMcollectiveCoord}
\bibinfo{author}{Tveten, E.~G.}, \bibinfo{author}{Qaiumzadeh, A.},
  \bibinfo{author}{Tretiakov, O.~A.} \& \bibinfo{author}{Brataas, A.}
\newblock \bibinfo{journal}{\bibinfo{title}{Staggered dynamics in
  antiferromagnets by collective coordinates}}.
\newblock {\emph{\JournalTitle{Phys. Rev. Lett.}}}
  \textbf{\bibinfo{volume}{110}}, \bibinfo{pages}{127208},
  \doiprefix\url{10.1103/PhysRevLett.110.127208} (\bibinfo{year}{2013}).

\bibitem{AFMdwWithSpinWaves}
\bibinfo{author}{Tveten, E.~G.}, \bibinfo{author}{Qaiumzadeh, A.} \&
  \bibinfo{author}{Brataas, A.}
\newblock \bibinfo{journal}{\bibinfo{title}{Antiferromagnetic domain wall
  motion induced by spin waves}}.
\newblock {\emph{\JournalTitle{Phys. Rev. Lett.}}}
  \textbf{\bibinfo{volume}{112}}, \bibinfo{pages}{147204},
  \doiprefix\url{10.1103/PhysRevLett.112.147204} (\bibinfo{year}{2014}).

\bibitem{PhysRevLett.111.217203}
\bibinfo{author}{Boulle, O.} \emph{et~al.}
\newblock \bibinfo{journal}{\bibinfo{title}{Domain wall tilting in the presence
  of the dzyaloshinskii-moriya interaction in out-of-plane magnetized magnetic
  nanotracks}}.
\newblock {\emph{\JournalTitle{Phys. Rev. Lett.}}}
  \textbf{\bibinfo{volume}{111}}, \bibinfo{pages}{217203},
  \doiprefix\url{10.1103/PhysRevLett.111.217203} (\bibinfo{year}{2013}).

\bibitem{SpinHallEffects}
\bibinfo{author}{Sinova, J.}, \bibinfo{author}{Valenzuela, S.~O.},
  \bibinfo{author}{Wunderlich, J.}, \bibinfo{author}{Back, C.~H.} \&
  \bibinfo{author}{Jungwirth, T.}
\newblock \bibinfo{journal}{\bibinfo{title}{Spin hall effects}}.
\newblock {\emph{\JournalTitle{Rev. Mod. Phys.}}}
  \textbf{\bibinfo{volume}{87}}, \bibinfo{pages}{1213--1260},
  \doiprefix\url{10.1103/RevModPhys.87.1213} (\bibinfo{year}{2015}).

\bibitem{PhysRevB.102.020408}
\bibinfo{author}{Reitz, D.}, \bibinfo{author}{Li, J.}, \bibinfo{author}{Yuan,
  W.}, \bibinfo{author}{Shi, J.} \& \bibinfo{author}{Tserkovnyak, Y.}
\newblock \bibinfo{journal}{\bibinfo{title}{Spin seebeck effect near the
  antiferromagnetic spin-flop transition}}.
\newblock {\emph{\JournalTitle{Phys. Rev. B}}} \textbf{\bibinfo{volume}{102}},
  \bibinfo{pages}{020408}, \doiprefix\url{10.1103/PhysRevB.102.020408}
  (\bibinfo{year}{2020}).

\bibitem{lebrun_long-distance_2020}
\bibinfo{author}{Lebrun, R.} \emph{et~al.}
\newblock \bibinfo{journal}{\bibinfo{title}{Long-distance spin-transport across
  the {Morin} phase transition up to room temperature in ultra-low damping
  single crystals of the antiferromagnet $\alpha$-{Fe2O3}}}.
\newblock {\emph{\JournalTitle{Nat. Commun.}}} \textbf{\bibinfo{volume}{11}},
  \bibinfo{pages}{6332}, \doiprefix\url{10.1038/s41467-020-20155-7}
  (\bibinfo{year}{2020}).

\bibitem{Slavin2023Artificialneurons}
\bibinfo{author}{Bradley, H.} \emph{et~al.}
\newblock \bibinfo{journal}{\bibinfo{title}{Artificial neurons based on
  antiferromagnetic auto-oscillators as a platform for neuromorphic
  computing}}.
\newblock {\emph{\JournalTitle{AIP Advances}}} \textbf{\bibinfo{volume}{13}},
  \bibinfo{pages}{015206}, \doiprefix\url{10.1063/5.0128530}
  (\bibinfo{year}{2023}).
\newblock \eprint{https://doi.org/10.1063/5.0128530}.

\bibitem{FundamentalNeuroscience}
\bibinfo{editor}{Squire, L.} \emph{et~al.} (eds.)
  \emph{\bibinfo{title}{Fundamental Neuroscience}}
  (\bibinfo{publisher}{Academic Press}, \bibinfo{address}{San Diego},
  \bibinfo{year}{2012}).

\bibitem{Bursting}
\bibinfo{author}{Overton, P.} \& \bibinfo{author}{Clark, D.}
\newblock \bibinfo{journal}{\bibinfo{title}{Burst firing in midbrain
  dopaminergic neurons}}.
\newblock {\emph{\JournalTitle{Brain research. Brain research reviews}}}
  \textbf{\bibinfo{volume}{25}}, \bibinfo{pages}{312—334},
  \doiprefix\url{10.1016/s0165-0173(97)00039-8} (\bibinfo{year}{1997}).

\bibitem{AFM-DWmagnonEmission}
\bibinfo{author}{Tatara, G.}, \bibinfo{author}{Akosa, C.~A.} \&
  \bibinfo{author}{Otxoa~de Zuazola, R.~M.}
\newblock \bibinfo{journal}{\bibinfo{title}{Magnon pair emission from a
  relativistic domain wall in antiferromagnets}}.
\newblock {\emph{\JournalTitle{Phys. Rev. Res.}}} \textbf{\bibinfo{volume}{2}},
  \bibinfo{pages}{043226}, \doiprefix\url{10.1103/PhysRevResearch.2.043226}
  (\bibinfo{year}{2020}).

\bibitem{FM-DWmagnonEmission}
\bibinfo{author}{Wieser, R.}, \bibinfo{author}{Vedmedenko, E.~Y.} \&
  \bibinfo{author}{Wiesendanger, R.}
\newblock \bibinfo{journal}{\bibinfo{title}{Domain wall motion damped by the
  emission of spin waves}}.
\newblock {\emph{\JournalTitle{Phys. Rev. B}}} \textbf{\bibinfo{volume}{81}},
  \bibinfo{pages}{024405}, \doiprefix\url{10.1103/PhysRevB.81.024405}
  (\bibinfo{year}{2010}).

\bibitem{Inhibition}
\bibinfo{editor}{Purves, D.} \emph{et~al.} (eds.)
  \emph{\bibinfo{title}{Neuroscience}} (\bibinfo{publisher}{Sinauer
  Associates}, \bibinfo{address}{Sunderland (MA)}, \bibinfo{year}{2001}).

\bibitem{inhibitionNegWeights}
\bibinfo{author}{Pfeiffer, M.} \& \bibinfo{author}{Pfeil, T.}
\newblock \bibinfo{journal}{\bibinfo{title}{Deep learning with spiking neurons:
  Opportunities and challenges}}.
\newblock {\emph{\JournalTitle{Frontiers in Neuroscience}}}
  \textbf{\bibinfo{volume}{12}}, \doiprefix\url{10.3389/fnins.2018.00774}
  (\bibinfo{year}{2018}).

\bibitem{PhysRevBShen}
\bibinfo{author}{Shen, L.} \emph{et~al.}
\newblock \bibinfo{journal}{\bibinfo{title}{Dynamics of the antiferromagnetic
  skyrmion induced by a magnetic anisotropy gradient}}.
\newblock {\emph{\JournalTitle{Phys. Rev. B}}} \textbf{\bibinfo{volume}{98}},
  \bibinfo{pages}{134448}, \doiprefix\url{10.1103/PhysRevB.98.134448}
  (\bibinfo{year}{2018}).

\bibitem{SulymenkoHematiteParams}
\bibinfo{author}{Sulymenko, O.~R.} \emph{et~al.}
\newblock \bibinfo{journal}{\bibinfo{title}{Terahertz-frequency spin hall
  auto-oscillator based on a canted antiferromagnet}}.
\newblock {\emph{\JournalTitle{Phys. Rev. Appl.}}}
  \textbf{\bibinfo{volume}{8}}, \bibinfo{pages}{064007},
  \doiprefix\url{10.1103/PhysRevApplied.8.064007} (\bibinfo{year}{2017}).

\end{thebibliography}

\section*{Acknowledgment}
This project has been supported by the Norwegian Financial Mechanism Project No. 2019/34/H/ST3/00515, ``2Dtronics''; and partially by the Research Council of Norway through its Centres of Excellence funding scheme, Project No. 262633, ``QuSpin''.

V. B. thanks Frank Mizrahi and Mark Stiles for inspiration and fruitful discussions.

\section*{Author contributions statement}
V.B. and J.A. conducted the simulations and analysis. S.L. provided technical support. A.Q. lead the project and discussions. All authors contributed to the manuscript. 

\section*{Additional information}
Additional simulation results can be found in the Appendix.

\newpage

\renewcommand{\thefigure}{S\arabic{figure}}

\renewcommand{\thetable}{S\arabic{table}}

\appendix
\noindent\huge\textbf{Appendix}\normalsize
\vspace{0.3cm}

\noindent We first demonstrate that the functionality of the proposed neuron is scalable. To prove that we use parameters of hematite, a prototype of two-sublattice AFM insulators.
Additionally, we show that in our proposed set up, the imaginary part of the spin mixing conductance is not relevant.

\section{Easy-plane hematite} \label{AppendixHematite}

\begin{figure}[b]
            \centering
            \includegraphics[width=0.5\linewidth]{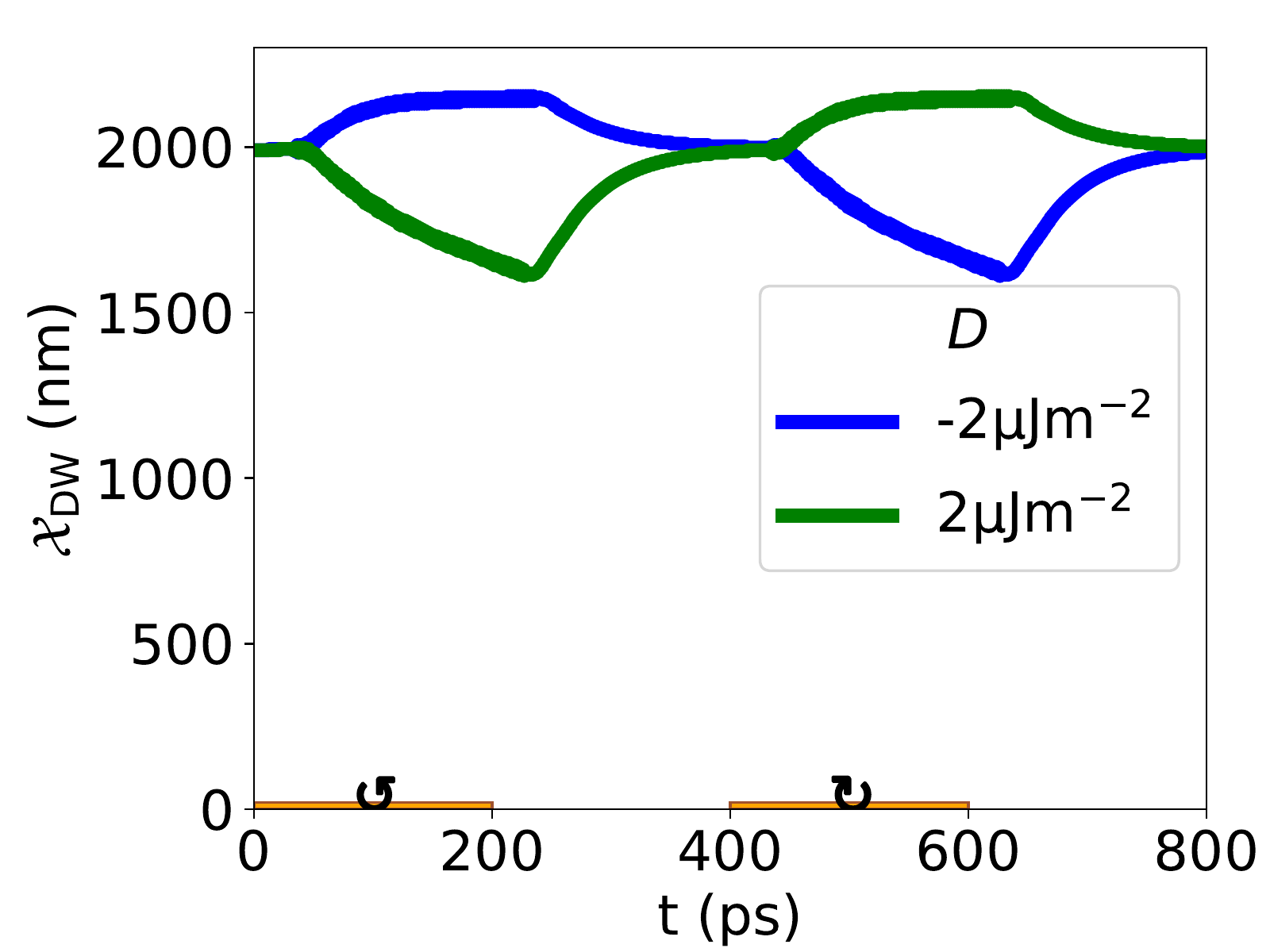}
            \caption{Magnetic field controlled DW motion in easy-plane hematite.}
            \label{fig:hematiteDWmotion}
        \end{figure}

We consider AFM hematite ($\alpha$-Fe$_3$O$_2$) above the Morin transition temperature where the system is in magnetic easy-plane phase.
In \cref{fig:hematiteDWmotion}, we present magnon induced domain wall (DW) motion for the easy-plane phase of hematite above the Morin transition, which is a prototype of orthorombic AFMs.

The motion is controlled by a magnetic field (position and duration indicated by orange area, to scale) with two opposite helicities (indicated by arrow). Two values of the bulk Dzyaloshinskii–Moriya interaction (DMI) $D$ are compared (blue vs green line). This is analogous to the magnetic field controlled motion presented in the main text with the four-stage protocol. Note that the system is larger compared to the toy model presented in the main article, due to the DW width. The DW equilibrium position is at \SI{2}{\micro\meter}.
        
As expected from our proposal based on our toy model parameters in the main text, both magnetic field helicity and direction (sign) of the DMI switch the direction of DW displacement. We choose an anisotropy profile $K(\vec{x}) = 10K_0\left[ \frac{1}{L_x} \left(x - \mathcal{X}_0\right)^2+ 1\right]$. 
Note that the slope of the profile can be tuned even larger as it was discussed in previous studies \cite{PhysRevBShen}. The simulation parameters for hematite \cite{SulymenkoHematiteParams}, are presented in \cref{tab:hematiteParams}.
        
        \begin{table}[b]
            \centering
            \caption{Simulation parameters for hematite \cite{SulymenkoHematiteParams}.}
             \begin{tabular}{@{}llll@{}} \hline
        		\textbf{Quantity}                & \textbf{Symbol}            & \textbf{Value}                & \textbf{Unit}                 \\ 
        		Length of AFMI layer	& $L_x$ & $\SI{3.0}{}$ & $\si{\micro \meter}$ \\
        		Width of AFMI layer	& $L_y$ & $\SI{20}{}$ & $\si{\nm}$ \\
        		Thickness of AFMI layer	& $L_z$ & $\SI{4}{}$ & $\si{\nano \meter} $\\
        		Grid size	& a & $\SI{4}{}$ & $\si{\nano \meter} $\\
        		Exchange stiffness	& $A_\text{AFM}$                      &  $\SI{76}{}$                    &         $\si{\femto\joule\per\meter}$ \\
        		Homogeneous exchange constant	& $A_h$                      &  $\SI{-460}{}$                    &         $\si{\kilo\joule \per \meter\cubed}$ \\
        		Easy-axis anisotropy constant & $K_\text{easy}$                      &    $-\SI{21}{}$                  &            $\si{\milli\joule \per \meter\cubed}$         \\
        		Hard-axis anisotropy constant & $K_\text{hard}$ 	 & $\SI{21}{}$ & $\si{\joule \per \meter\cubed}$ \\
        		Saturation magnetization & $M_s$                      &      $\SI{2.1}{}$                  &                 $\si{\kilo\ampere\per \meter}$ \\ 
        		Gilbert damping&           $\alpha$           &   $\SI{0.0003}{}$                    &                      1\\ 
        		Homogenous DMI coefficient&           $D_h$            &        $\SI{4.6}{}$             &                    $\si{\kilo\joule \per \meter\cubed}$  \\
        		Time step &           $\Delta t$            &     $\SI{2}{}$     &                    $\si{\femto\second}$  \\
        	\end{tabular}\\
            \label{tab:hematiteParams}
        \end{table}
       
     In summary, our proposed neuron can be realized in AFM systems with generic orthorhombic symmetry. Excitation timescales should be tuned for each chosen material. 
     
    \section{Contribution of the  imaginary part of the spin mixing conductance} \label{appendixImPart}
    In general, the imaginary part of the spin mixing conductance is dependent on the quality of the interface between the heavy metal layer and the magnetic layer. This term is negligible for dirty interfaces.
    The spin pumping has the following general form \cite{ArneSpinMixingConductance,ArneSpinPumping}
    \begin{equation} \centering \label{eq:spinAccum}
    \bm{\mu}(t) := G_r^{\uparrow \downarrow} \big(\bm{n}(t,\bm{r})\times \dot{\bm{n}}(t,\bm{r})+\bm{m}(t,\bm{r})\times \dot{\bm{m}}(t,\bm{r})\big)- G_i^{\uparrow \downarrow}\dot{\bm{m}}(t,\bm{r}) ,
    \end{equation}
    with the N\'eel vector $\bm{n}=\frac{\bm{m}_A-\bm{m}_B}{2}$ and magnetization $\bm{m}=\frac{\bm{m}_A+\bm{m}_B}{2}$, where $G_r^{\uparrow \downarrow}$ and $G_i^{\uparrow \downarrow}$ are the real part and  the imaginary part of the spin mixing conductance, respectively.
    
\begin{figure}
        \centering
        \includegraphics[width=0.5\linewidth]{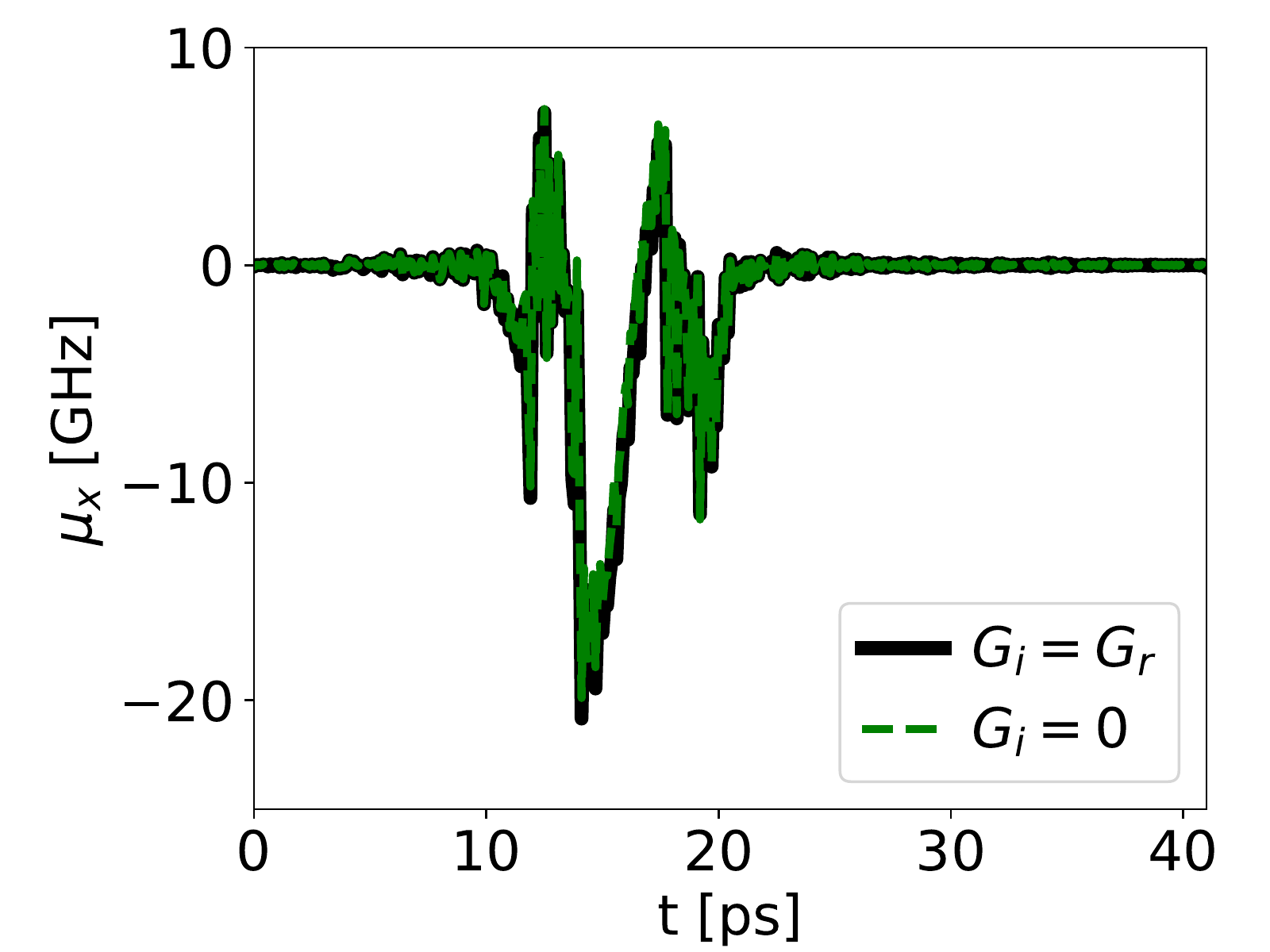}
        \caption{Comparison of read out spin pumping signal including and not including the imaginary spin mixing conductance.}
        \label{fig:ralImaginary}
    \end{figure}

    In order to check the qualitative and quantitative effects of including $G_i^{\uparrow \downarrow}$, we compare two extreme cases, i.e., $G_i^{\uparrow \downarrow}=G_r^{\uparrow \downarrow}$ (large imaginary part) and $G_i^{\uparrow \downarrow}=0$ (zero imaginary part). As shown in \cref{fig:ralImaginary}, both read outs are the same, suggesting that the imaginary part of the spin mixing conductance can be neglected in our set up geometry.
    
\end{document}